\documentclass[fleqn,usenatbib]{mnras}

\usepackage{newtxtext,newtxmath}

\usepackage[T1]{fontenc}

\DeclareRobustCommand{\VAN}[3]{#2}
\let\VANthebibliography\thebibliography
\def\thebibliography{\DeclareRobustCommand{\VAN}[3]{##3}\VANthebibliography}

\usepackage{ae,aecompl}

\usepackage{graphicx}	
\usepackage{amsmath}	
\usepackage{aas_macros}

\newcommand{\Msun}{M$_\odot$}

\newcommand{\degree}{\ensuremath{^\circ}}

\title[Dynamical modelling PGC 046832]{Dynamical modelling of the twisted galaxy PGC 046832}

\author[M. den Brok et al.]{
  Mark den Brok,$^{1}$\thanks{E-mail: mdbrok@aip.de}, Davor Krajnovi{\'c},$^{1}$ Eric Emsellem,$^{2}$ Jarle Brinchmann,$^{3,4}$ Michael Maseda$^{3}$\\
$^1$Leibniz-Institut f\"ur Astrophysik Potsdam (AIP), An der Sternwarte 16, 14482, Potsdam, Germany\\
$^2$European Southern Observatory, Karl-Schwarzschild-Str. 2, D-85748 Garching, Germany\\
$^3$Leiden Observatory, Leiden University, PO Box 9513, NL-2300 RA Leiden, the Netherlands\\
$^4$Instituto de Astrof\'isica e Ci{\^e}ncias do Espa\c{c}o, Universidade do Porto, CAUP, Rua das Estrelas, PT4150-762 Porto, Portugal\\
}

\date{Accepted XXX. Received YYY; in original form ZZZ}

\pubyear{2021}

\begin{document}
\label{firstpage}
\pagerange{\pageref{firstpage}--\pageref{lastpage}}
\maketitle

\begin{abstract}
   We analyse VLT/MUSE observations of PGC~046832, the  brightest cluster galaxy of Abell 3556. The velocity structure of this galaxy is startling and shows two reversals in sign along the minor axis, and one along the major axis.
  We use triaxial Schwarzschild models to infer the intrinsic shape, central black hole mass and orbit distribution of this galaxy. The shape determination suggests that the galaxy is highly triaxial in the centre (almost prolate) but has a low triaxiality (almost oblate) in the outer parts.
  The orbit distribution of the best-fit Schwarzschild model shows that the kinematic reversal along the projected minor axis is driven by a slight asymmetry in the distribution and amount of long axis tubes in the inner parts. The kinematic reversals along the projected major axis are driven by a high fraction of counter-rotating orbits at intermediate radii in the galaxy. Using chemical tagging of orbits in the Schwarzschild model, we do not find evidence for any association of these orbits with specific stellar population parameters. Although the inner part of the galaxy almost certainty formed through one or more dry mergers producing the prolate shape, the outer parts are consistent with both  accretion and in situ formation. While axisymmetric models suggests the presence of a supermassive black hole with mass $\sim 6 \times 10^9$\Msun and $\sim 10^{10}$\Msun (with Schwarzschild and Jeans modelling, resp), triaxial Schwarzschild models provide only an upper limit of $\sim 2 \times 10^9$\Msun.
\end{abstract}

\begin{keywords}
galaxies: elliptical and lenticular, cD -- galaxies: kinematics and dynamics 
\end{keywords}



\section{Introduction}\label{sec:intro}
The most massive early-type galaxies are structurally and kinematically different from their lower-mass counterparts. Photometric analyses have shown that these galaxies are on average rounder in projection \citep[e.g.][]{TreMer96,EmsCapKra07} and have isophotes that are more boxy \citep{NieBen89}. Their central surface brightness profiles often deviate from perfect S\'ersic models. Contrary to lower-mass early-type galaxies, which often show an enhanced central surface brightness, the most massive galaxies show often a deficit of light at their centre and are therefore called {\it cored} \citep[e.g.][]{Kor85,NieBenSur91}.
The amount of ordered motion, characterized by $v/\sigma$, has been found to be lower in the most massive galaxies than in lower-mass ellipticals \citep{BerCap75,Ill77}. Using $\lambda_R$ as a more robust proxy for specific angular momentum, \citet{EmsCapKra07} divided early-type galaxies galaxies in slow- and fast rotators and showed that fraction of slow rotators appears to be higher for the highest mass galaxies. Employing a bigger sample of early-type galaxies, \citet{EmsCapKra11} showed that slow rotaters have in general dynamical masses above $10^{10.5}$ \Msun.

The formation and internal structure of the most massive galaxies is not yet
fully understood. It is established that most their stars were formed during
intense bursts at high redshift \citep[e.g.][]{ThoMarBen05,GreJanMa15}, which
simulations show to be intertwined with and or followed by one or more major
mergers. However, it is still not known what fraction of stars seen today in
the most massive galaxies actually formed in the main predecessor, and how much
of their mass was acquired during mergers. Simulations show that the accreted mass can be higher than the in-situ formed mass \citep[e.g.][]{PilNelHer18}.

Clues to the formation of the most massive galaxies can be obtained by studying their intrinsic shapes and orbit distributions. The intrinsic shape of a major dry merger remnant  is dependent on the mass ratio, gas fractions and orbital parameters of the merging galaxies \citep[e.g.]{JesNaaPel07,LiMaoEms18}. Similarly, the fractions of short axis tubes and box orbits may provide clues to the mass ratios of the progenitor galaxies \citep[e.g.][]{JesNaaBur05}.

The misalignment of the angular momentum vector at different radii points at triaxiality. \citet{Bin85} and \citep{FraIlldeZ91} used this to find the distribution of intrinsic shapes of samples of elliptical galaxies. \citet{TenBusLon93}, \citet{Sta94, Sta94a}, \citet{Sta01} and \citet{StaEmsPel04} used simple dynamical models to constrain the shape of elliptical galaxies. A major step forward has been the development of self-consistent triaxial orbit superposition models by   \citet{vanvanVer08}, who also showed how the triaxiality of NGC 4365 varies with radius. 

We have performed a survey with VLT/MUSE to obtain integral field spectroscopy of the most massive galaxies in the Universe in the densest environments.  The sample, which was presented in \citet{KraEmsden18}, consists of 25 galaxies, 14 of which were in the Shapley Super Cluster (SSC), and 11 BCGs outside the SSC. One of the galaxies in the sample -- PGC 046832, the BCG of Abell 3556, one of the galaxy clusters in the SSC -- has a rather unique velocity structure, which has triggered the present study. The velocity map of this galaxy shows 1) an inner KDC (2\arcsec in size), 2) two more reversals of the ``observed'' angular momentum, and 3) rotation around the major axis. These properties of the velocity field make this galaxy a very rare object \citep[see e.g.][]{KraEmsCap11}. \citet{RanJohNaa18} and \citet{RanJohNaa19} show that such reversals in rotation direction may be a consequence of the inspiral of two supermassive black holes in a major merger. This scenario would require a very high mass for the final merged black hole. The black hole mass might therefore provide an additional clue to the formation process of this particular galaxy.

In this paper we model this galaxy with Schwarzschild's superposition method. This method has been employed to, among other things, study the orbit distribution in galaxies \citep{vandeZvan08,Zhuvanvan18}, and constrain the characteristics of their dark matter halos \citep[e.g.][]{ThoSagBen07}.

Schwarzschild models have a long history of being used for determining black hole masses at galaxy centres. \citet{RicTre85} and \citet{DreRic88} showed, using spherically symmetric Schwarzschild models, that the centres of  M87, M31 and M32 provide evidence for black holes. The extension of Schwarzschild models to axially symmetric potentials has been used widely to determine black hole masses for axisymmetric systems \citep[e.g.][]{vanCredeZ98,GebRicKor00,VerCapCop02,CapVervan02,GebRicTre03,KraMcDCap09,GebTho09,SchGeb11,RusThoSag13,KraCapMcD18,ThaKraCap19}. A Schwarzschild code  with triaxial symmetry was published by \citet{vanvanVer08}, which has been used for modelling black holes in axially symmetric and triaxial systems \citep[e.g.][]{vandeZ10,vanGebGul12,SetvanMie14,LieQueMa20}.

In Section \ref{sec:data} we present the integral field spectroscopy and photometry data used in this paper.  We derive the kinematics in stellar population parameters in Sec. \ref{sec:analysis}. We then proceed to model the kinematics with orbit superposition models in Sec. \ref{sec:shape} to infer the intrinsic shape and dark matter content of the galaxy. We derive and compare a black hole mass with Jeans models and orbit superposition models in Sec \ref{sec:mbh}. Stellar population parameters are derived in \ref{sec:ssp}. 

Throughout the paper we use the following cosmological parameters: ($\Omega_M$, $\Omega_{\Lambda}$, h) = (0.315, 0.685, 0.674) \citep{Pla18}, which for PGC 046832's redshift of $z= 0.048$ corresponds to an angular diameter distance of 201 Mpc.

\section{Data}
\label{sec:data}
\subsection{VLT/MUSE integral field spectroscopy}
We observed PGC 046832 with VLT/MUSE as part of the MUSE Most Massive Galaxies (M3G) survey in MUSE GTO time. The initial 3 observation series were carried out in 2015 and 2016 and consisted of 4$\times $630s exposures combined with an additional 4$\times$180s exposures (Table \ref{tab:summary_muse_obs}). We
further re-observed the galaxy using the ground-layer adaptive optics
corrections with the hope of achieving a higher spatial resolution.
The data were reduced with version 2.6 of the MUSE data reduction pipeline
\citep{WeiPalStr20}. Steps in the data reduction processes included bias
subtraction, flat fielding and wavelength calibration. Sky was observed in
separate sky fields and subtracted from the science exposures using the Zurich Atmospheric Purge \citep[ZAP;][]{SotLilBac16}.
We used the geometry and
astrometry tables produced during each GTO observing run. To benefit of the
higher spatial resolution of the AO data, we produced two different merged data cubes: a high resolution cube based exclusively on the AO observing run (38 minutes on source
time), and a deeper cube with somewhat
lower resolution spatial resolution based on all observations (1.5 hr on source time), both with a field-of-view of 60\arcsec$\times$60\arcsec, a spatial sampling of 0\farcs20 and a spectral sampling of 1.25\AA. 
\begin{table}
\caption{Summary of MUSE observations. The spatial resolution was measured at 5500\AA.}\label{tab:summary_muse_obs}
\begin{tabular}{ccc}
\hline
Date & Exposure time & Resolution \\
 & [s] & FWHM [arcsec]\\
\hline
22-05-2015 & 4 $\times$ 180 & 0.82\\
11-03-2016 & 2 $\times$ 630 & 1.4\\
12-03-2016 & 2 $\times$ 630 & 0.85\\
04-03-2019 & 4 $\times$ 570 & 0.67\\
\end{tabular}
\end{table}

\subsection{Photometry}
\subsubsection{HST/WFPC2}
PGC 046832 was observed with HST/WFPC2 as part of program 8683 (PI: van der Marel) in the F814W band, in two exposures of 500s each. These imaging data are important to constrain the luminosity profile of the galaxy in the inner parts. We downloaded the reduced data through the Hubble Legacy Archive. The data have a pixel scale of 0\farcs05 pixel$^{-1}$.  To derive the PSF for these data we generate an artificial PSF with {\it tinytim} \citep{KriHooSto11}. For the parametrization of the light profile we need this PSF expanded as a series of concentric Gaussians.
We therefore subsequently model the model PSF as 4 Gaussians using \textsc{galfit} \citep{PenHoImp10}. 

\subsubsection{VST/OmegaCam}
PGC046832 was observed  with VST/OmegaCam on Cerro Paranal in the {\it g, r}
and {\it i} bands (PI Merluzzi) for 1400s, 2600s and 1000s.  We received the reduced images from the PI. The data reduction procedures are described in \citet{MerMerBus15}. The pixel size of the images is 0\farcs21 pixel$^{-1}$. We measured the seeing by fitting Gaussian models to foreground stars in the field of view. The seeing varies between 0\farcs6 in the {\it r} band and 0\farcs72 in the {\it i} band.

\section{Analysis}
\label{sec:analysis}

\subsection{Structural parameters}
In order to have an accurate measurement of the core size, we carry out a fit to the
HST and VST photometry of PGC 046832 to determine the structural parameters.  We do
this by fitting the HST and VST data in parallel, with a custom 
code.
\citet{LaivanLau03} found a core size of $R_b = 0\farcs47$ by fitting Nuker models to the HST data. 
As Nuker models have been shown to be affected by the radial range over which
they are fitted \citep{GraErwTru03,GraErwTru04}, we choose instead a combination of a
S\'ersic \citep{Ser68} profile and a core-S\'ersic profile to describe the
galaxy. The latter profile is given by 
\begin{eqnarray}
 I(R) = I' \left[ 1 + \left(\frac{R_b}{R}\right)^{\alpha}\right]^{\frac{\gamma}{\alpha}} \exp\left[
  -b_n \left( \frac{R^{\alpha} + R_b^{\alpha}}{R_e^{\alpha}}
  \right)^{1/\alpha n}\right],   
\end{eqnarray}
whereas the former profile can be obtained through the limit 
$R_b \rightarrow 0$. Like in \texttt{galfit} \citep{PenHoImp10}, we allow for
boxiness through the parameter $C$ that changes the the radial coordinate into
generalized ellipses \citep[e.g.][]{AthMorWoz90}:
\begin{eqnarray}
R = \left[ x^{2+C} + \left(\frac{y}{q}\right)^{2+C} \right]^{1/(2+C)}.
  \end{eqnarray}
Each component has a free
ellipticity, position angle (P.A.) and shape parameters, while sharing the central
position. We subsample each model pixel by a factor 4 in the outer parts, and
a factor 100 in the center, rebinning them and convolving them with the
PSF. We define a likelihood through $\log(\mathcal{L}) = -\frac{1}{2} \left(
  \chi_{\mathrm{HST}}^2 + \chi_{\mathrm{VST}}^2 \right)$. The best fit
  parameters are given in Table \ref{tab:structfitpar}. During the fit, we
  allow for a difference in zeropoint between the two datasets. The magnitude
  of this difference, 0.05 mag, can probably be fully attributed to
  uncertainties in the conversion between photometric systems and slightly
  different filter curves. The magnitude of the core-S\'ersic profile is the magnitude of the S\'ersic model with the same parameters but $R_b \rightarrow 0$, i.e. we do not compensate the magnitude for the missing light in the core. The core radius $R_b = 0\farcs41 \pm 0\farcs03$ is slightly lower than the core radius of 0\farcs47 determined by \citet{LaivanLau03}, either because of the additional constraint on the light profile from ground-based data or due the use of a core-S\'ersic profile instead of a Nuker profile.  

\begin{table*}
\caption{Summary of structural fits. Note that the magnitude of the
  core-S\'ersic model is the magnitude of the equivalent model with $R_b=0$.}\label{tab:structfitpar}
\begin{tabular}{cccccccccc}
\hline
Model & F$_{\nu}$  & $R_{e}$   &   n  & $\epsilon$ &  P.A. & C &  $R_b$     & $\gamma$ & $\alpha$ \\  
      & [Vega mag] & [arcsec] &      &  & [deg] &  &  [arcsec] &  \\
\hline
\\
\vspace{2mm}
Core-S\'ersic & 13.2$_{-0.02}^{+0.02}$ & 6.39$_{-0.15}^{+0.19}$ &
  5.13$_{-0.05}^{+0.05}$ & 0.38$_{-0.004}^{+0.004}$ & $124.8_{-0.03}^{+0.04}$ &
  $0.21_{-0.01}^{+0.01}$ & $0.41_{-0.03}^{+0.03}$ & $0.43_{-0.01}^{+0.01}$ &  $6.7_{-0.2}^{+0.2}$    \\
S\'ersic  & $12.79_{-0.01}^{+0.01}$ & $17.36_{-0.04}^{+0.04}$ &
$1.53_{-0.01}^{+0.01}$ &  $0.27_{-0.003}^{+0.003}$  &  $116.5_{-0.2}^{+0.2}$ &
$0.29_{-0.01}^{+0.01}$ \\
\end{tabular}
\end{table*}

\begin{figure}
  \includegraphics[angle=0, width=.50\textwidth]{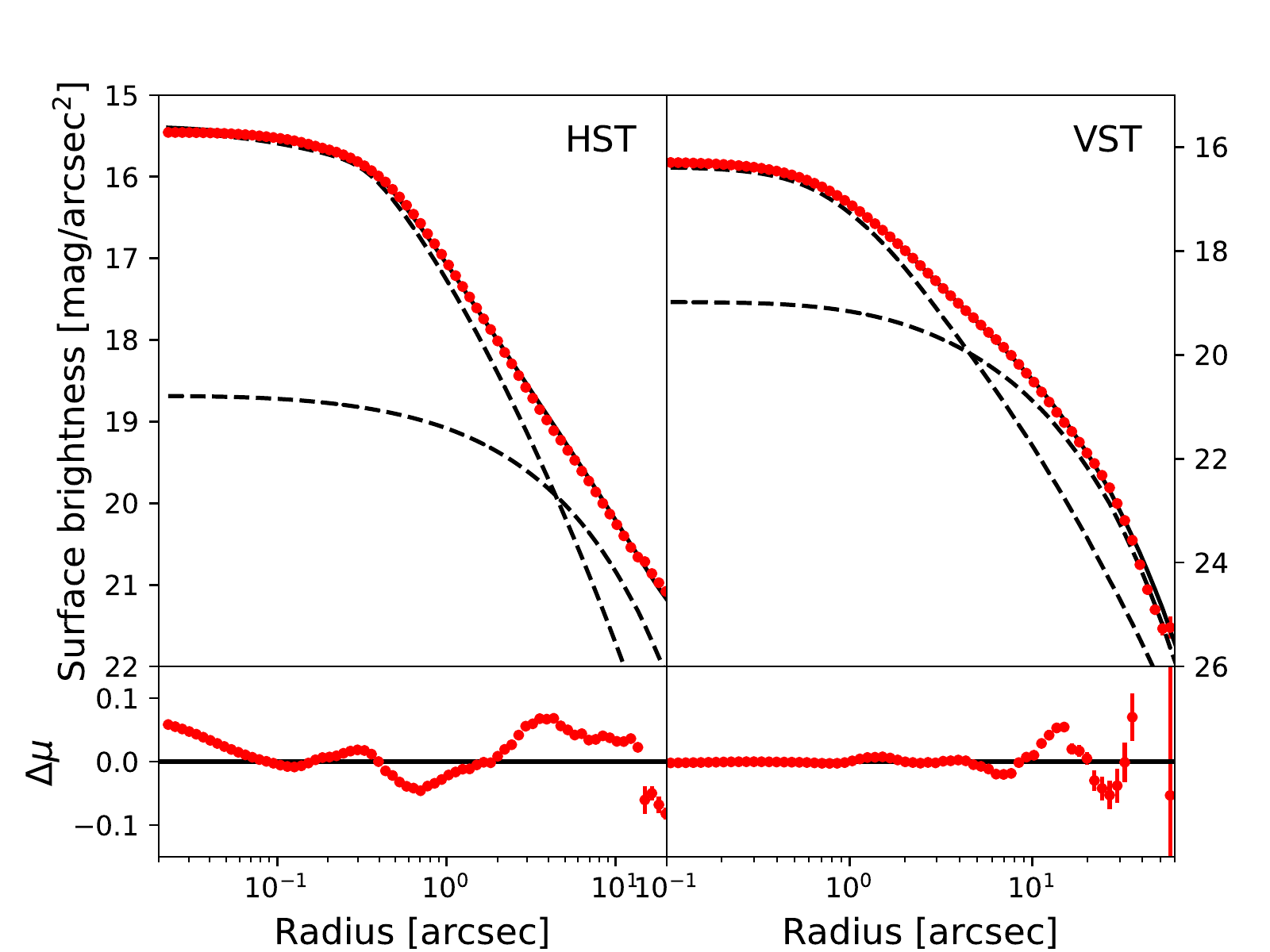} 
  \caption{One-dimensional summary of the parallel structural parameter fit to
  the HST data (left) and VST data (right). Data are shown as red dots, the
  models with a solid black line. The individual (PSF-convolved) core-S\'ersic and S\'ersic
  models are shown as dasked lines. Lower panels show the surface brightness residuals of {\it data -- model}.}\label{fig:structpars}
\end{figure}

\subsection{Derivation of the luminous MGE}\label{sec:analysis_mge}
\begin{figure}
  \includegraphics[angle=0, width=.48\textwidth]{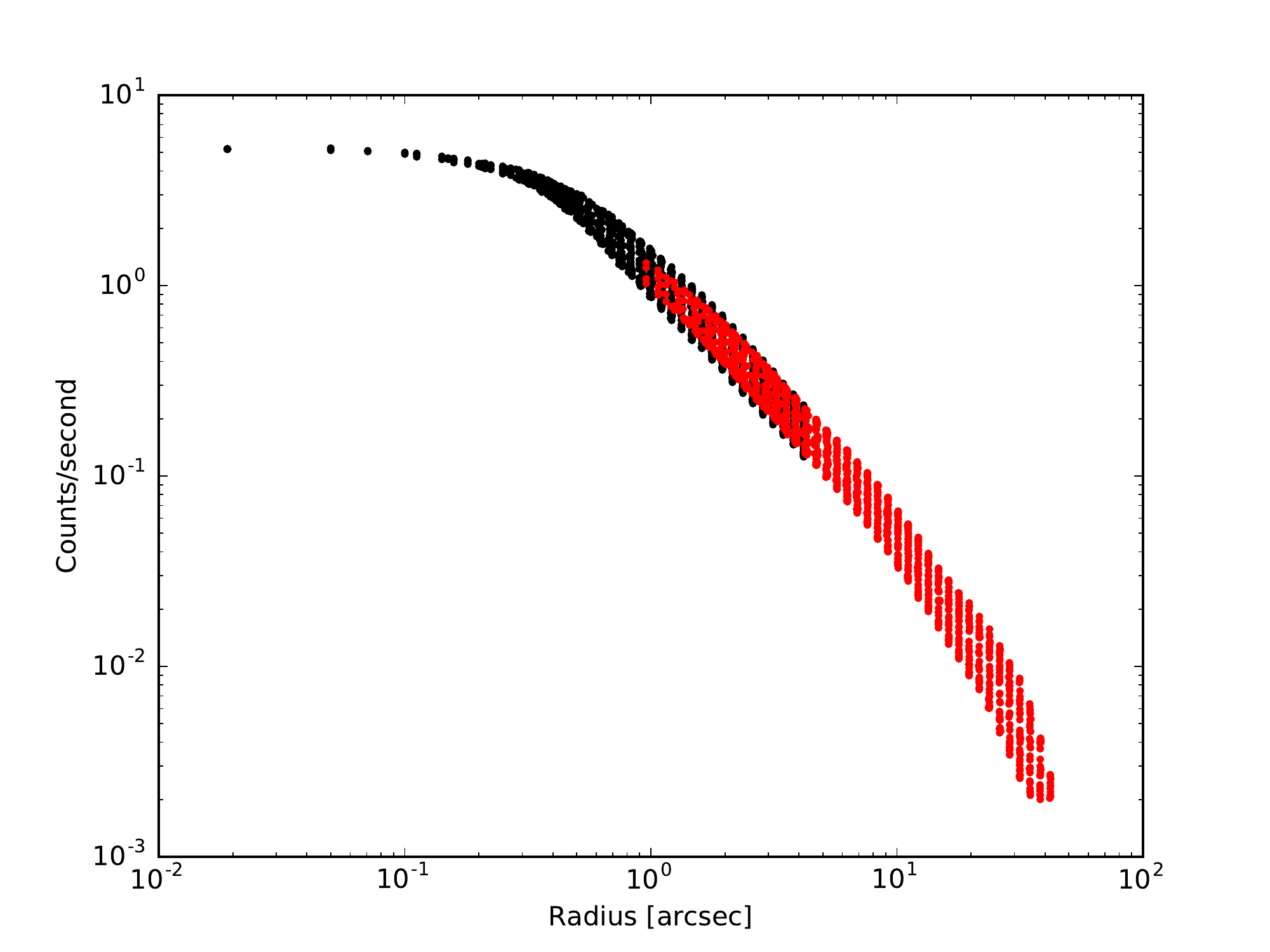}
  \caption{Surface brightness in units of counts/second/pixel for the PC2 image (black dots) and the OmegaCAM image (red points). The brightness units from the OmegaCAM image were converted to the PC2 pixel size and zeropoint. }\label{fig:mge_sectors}
  \end{figure}

To create a combined MGE of the two different imaging datasets, we use the
\textsc{sectors\_photometry} and \textsc{mge\_fit\_sectors} codes from
\citet{Cap02}. We convert the calibrated OmegaCAM image to the counts per
pixel system of the PC2 data by multiplying with the appropriate difference in
zeropoints, gain, exposure time and pixel size. We show the photometry
measurements as a function of radius in in Fig. \ref{fig:mge_sectors}
We use \textsc{SExtractor} \citep{BerArn96} in
standard configuration to identify bright sources in the OmegaCAM and HST
images, which we mask when we derive the photometry per sector. We then
provide the photometry per sector of both images to the
\textsc{mge\_fit\_sectors} code to derive the MGE. We derive two separate
MGEs: one including photometric twists and one without, the former for
modeling the galaxy as a triaxial ellipsoid, the latter to model it as an
axisymmetric system.

\begin{figure*}
  \includegraphics[angle=0, width=1.\textwidth]{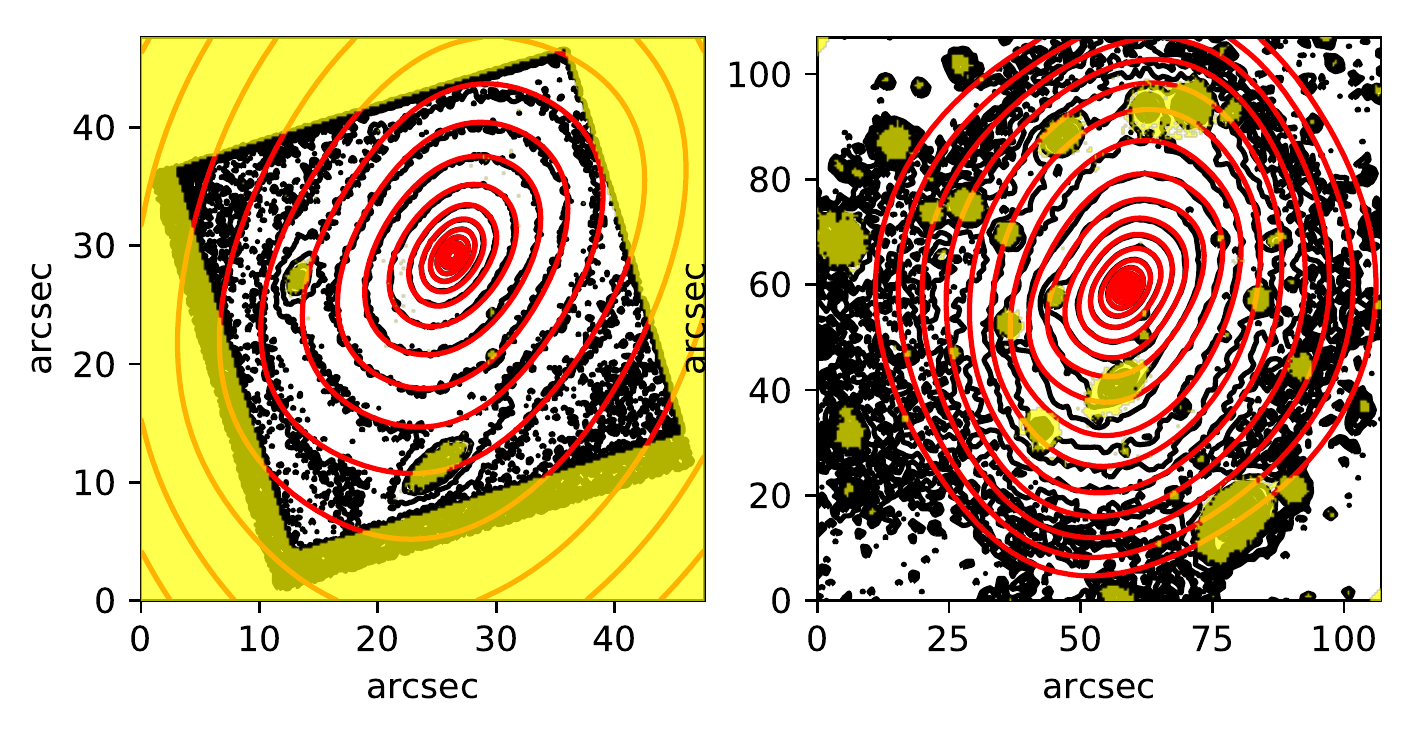}
  \caption{Contours describing the results of the simultaneous MGE fit to the
    PC2 data and OmegaCAM data. Red contours are the MGE. Yellow areas were
    not included in the MGE fit.}\label{fig:mge_contours}
  \end{figure*}

To allow the largest range in viewing angles for the dynamical modelling, we 
enforce a roundness constraint on the MGE components of  $q' > 0.62$, corresponding to the lowest value observed in the VST photometry.
We found that leaving the position angle (PA) $\Psi_i$ of each MGE component completely free leads to a PA that oscillates with radius, and often to pairs of MGE components with similar width, but different PA, possibly to compensate for the boxiness of the galaxy. We therefore added a regularization term of the form
$ R = \lambda \sum_{i=2}^N (\Psi_{i-1} - \Psi_{i})^2$ to the merit function in
 \textsc{mge\_fit\_sectors}. This regularization term suppresses the oscillatory behaviour while still reproducing the photometric twist seen at larger radii. 
 We subsequently correct the MGE for foreground extinction of A$_i \approx 0.098$ mag \citep{SchFin11}, cosmological surface brightness dimming (a factor 1.2) and a K correction of 0.04 mag \citep{ChiMelZol10}. This corrected MGE is tabulated in Table \ref{tab:MGE_twists}, whereas we show the axisymmetric
 MGE in the Appendix in Table \ref{tab:MGE_notwists}.
 The contours of the MGE, convolved to the resolution of the WF/PC2 and VST images, are shown in Fig. \ref{fig:mge_contours}.   

\begin{table}
\caption{MGE of PGC 046832 allowing photometric twists. The surface brightness has been corrected for foreground extinction, dimming and K-correction.}
\label{tab:MGE_twists}  
\begin{tabular}{ccccc} 
\hline
No. & SB$_0$  & $\sigma$  & $q'$ & $\Psi_i$\\ 
 & [L$_{\odot}$/pc$^2$] & [arcsec] &  & [$\deg$]\\ 
\hline
1 &  2054.6  &   0.079  &   0.84  &   0.0 \\
2 &  5484.3  &   0.285  &   0.85   &  0.0  \\
3 &  7567.0  &   0.554  &   0.66  &   0.0  \\
4 &  2702.6  &   1.227  &   0.62  &   0.0  \\
5 &  1235.5  &   2.363 &    0.65  &   0.0  \\
6 &   435.6   &  5.315  &   0.69  &   0.0  \\
7 &   157.3  &   10.621  &   0.64  &   3.7 \\
8 &   65.2  &   21.082 &    0.76   &  14.0 \\
\end{tabular}
\end{table}

\subsection{Derivation of the kinematics}
In order to achieve high-enough signal-to-noise (S/N) to derive the kinematics
from the MUSE spectra, we use the Voronoi binning code of \citet{CapCop03}.
To combine both high and low-resolution data sets, we run the Voronoi code twice. We first run the code on the high resolution data set with the outer parts masked, and then run the Voronoi code on the full data set with the bins from the first run masked. The sets of bins are then combined to form one dataset.

We calculate the average S/N for each
spaxel at the wavelength range 5250--5350 \AA (rest frame), which is
relatively devoid of strong absorption lines. We bin to achieve a S/N of 60\AA$^{-1}$, using the formal uncertainties derived from the data reduction. 

\begin{figure*}
  \includegraphics[angle=0, width=\textwidth]{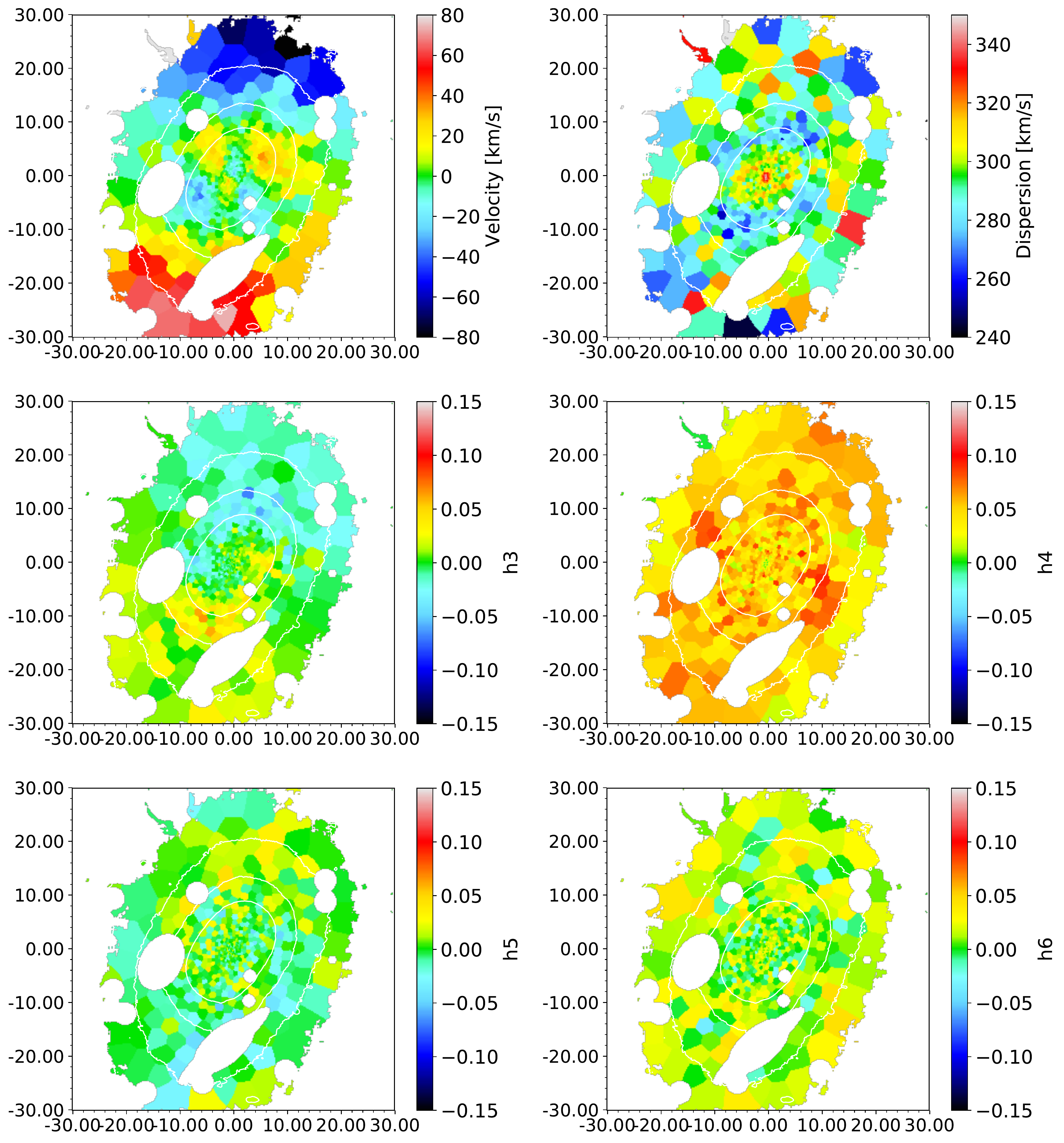} 
  \caption{Kinematic data of PGC 046832 derived using pPXF and MILES stars. Upper panels from left to right: velocity, velocity dispersion. Second row:  $h_3$ and $h_4$. Bottom row $h_5$ and $h_6$. }\label{fig:datakin_full}
\end{figure*}

\begin{figure*}
  \includegraphics[angle=0, width=\textwidth]{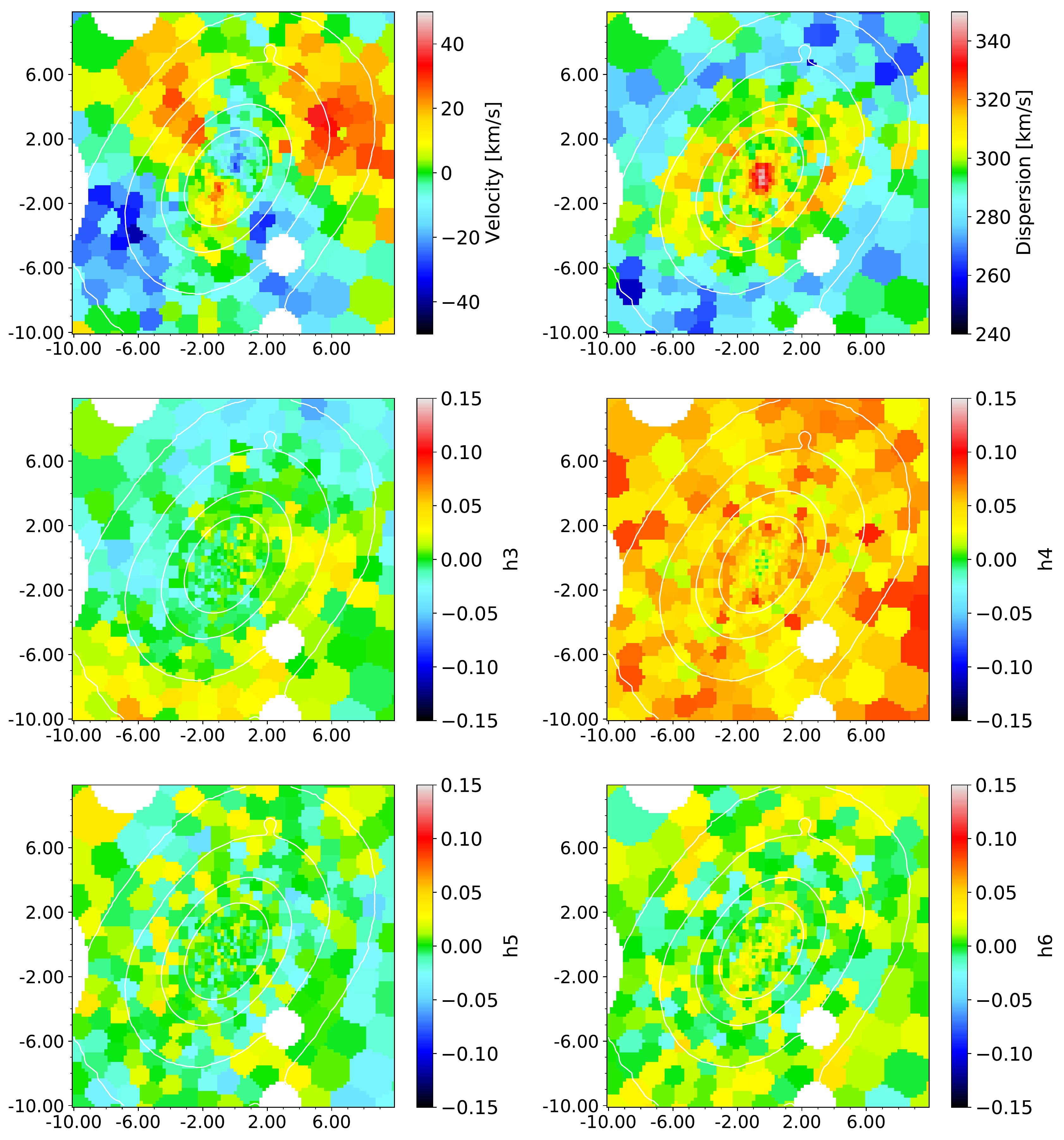} 
  \caption{Central zoom in of the kinematic data. Panels as in Fig. \ref{fig:datakin_full}.  }\label{fig:datakin_full_sub}
\end{figure*}

For each Voronoi bin, we determine the kinematics using \texttt{ppxf}
\citep{Cap17} and MILES stellar spectra \citep{SanPelJim06} over the wavelength range 4800-7100\AA. We allow for an additive polynomial up
to order 8, and Gauss-Hermite coefficients \citep{vanFra93} up to order 6. We mask bright sky lines, telluric bands, the wavelength gap caused by the laser, and -- even though ionized gas is not detected in the MUSE data \citep{PagKraden20} -- the centres of H${\alpha}$ and H${\beta}$. After fitting we a posteriori estimate the S/N of each Voronoi bin by determining
the standard deviation of the residual spectrum after subtraction of the best
fitting template and the average signal in the best-fit template. 
We then carry out a Monte Carlo simulation for each bin, during which we
perturb 100 times the spectrum of each bin by the residual noise, and refit each
simulated spectrum with \texttt{ppxf} with the same parameters as before. We
then use as best fit and uncertainty of each kinematic quantity the average
and 1-sigma deviation from the simulation. Typical uncertainties on $v$ and $\sigma$ are about 10 km/s and 0.02 on Gauss-Hermite coefficients.  The kinematics
are shown in Fig. \ref{fig:datakin_full}. Because the kinematic data are particularly rich at smaller scales we also show a zoom in of the centre of the kinematic data in Fig. \ref{fig:datakin_full_sub}.

\section{Dynamical modeling}
\label{sec:modeling}
\begin{table*}\caption{Summary of triaxial Schwarzschild modeling procedures. Note that the DM Halo mass denotes the virial mass in terms of the total MGE mass.}
\begin{tabular}{lllll}
Method & DM Halo & M$_{BH}$ & $\theta, \phi, \Psi$ & M/L\\
\hline
1) Grid search & Mass=$1000,1500,2000$ & $2\times10^9$ \Msun & Sampled through ($p,q,u$) & Grid\\
2) Ellipsoidal search & Mass free & Free ($> 10^8$ \Msun) & Free & Free\\
3) Ellipsoidal search increased \# orbits & Mass free & Free ($> 10^8$ \Msun) & Free & Free\\
4) BH fit & Mass=1475  & Grid  & (61$^{\circ}$, -59$^{\circ}$, 75$^{\circ}$) & Grid \\
5) Effective \# free parameters & Mass=1500 &  $2\times10^9$ \Msun& Limited grid in ($p,q,u$) & Grid\\
\end{tabular}
\end{table*}

\subsection{Intrinsic shape}\label{sec:shape}
Dynamical models of a galaxy can provide constraints on the viewing angles under which a galaxy is seen, as these models simultaneously predict where stars are in a galaxy and how these stars move. Some viewing angles will lead to intrinsic mass and light distributions that cannot reproduce the observed kinematics of galaxies. For oblate galaxies it can already be difficult to recover the inclination angle. In this case the inclusion of higher order Gauss-Hermite terms in the fit of the dynamical model to the kinematics can help constraining the inclination \citep{KraCapEms05}. For a triaxial galaxy, the recovery of the intrinsic shape is even more complicated because it requires constraining three viewing angles instead of one. In this case the presence of kinematically decoupled
components and isophotal twists may still provide a way to break degerenacies between viewing angles \citep{vanvan09}.

We use the triaxial Schwarzschild code of \citet{vanvanVer08} to model PGC 046832. Constraining the shape requires sampling all three viewing angles ($\theta, \phi, \Psi$). 
In Section \ref{sec:analysis_mge}, we decomposed the light profile of the
galaxy into a series of concentric Gaussians. Each of these Gaussians has an
observed axis ratio $q_j'$ and misalignment angle $\Psi_j$. Given the three viewing angles, the observed axis ratio and the misalignment angle, the Schwarzschild code deprojects each Gaussian following the equations of \citet{MonBacEms92}. We do note that the deprojection is not unique, and that other solutions may exist.

For each deprojected mass distribution, the Schwarzschild code generates two orbit libraries: one consisting mostly of short and long axis tube orbits and a small amount of box orbits, and the other one consisting mainly of box orbits and a small amount of tube orbits \citep{vanvan09}. Each orbit library consists of $n_E \times n_{\theta} \times n_{R} =n_E \times n_{\theta} \times n_{phi} = 31\times6\times6$ different starting positions (for the regular orbit library and the box orbit library) with 5 different dithers for each orbit integral.

For each position along a (PSF convolved) orbit, a velocity histogram along the line-of-sight is calculated and stored in the orbit library. The orbit libraries are then used as input to a non-linear least squares fit of the velocity, dispersion and Gauss-Hermite terms, in our case up to $h_6$ to the kinematic data, while simultaneously constraining the projected light and deprojected mass distribution within a few percent.  

We search for the best-fit model using a combination of two techniques: a grid-based search, and an algorithm that given the already calculated models explores the parameter space where a better fitting model could be found.
The first technique we employ is a grid-based search. Although one could sample the viewing angle space linearly,  \citet{vanvan09} advise to instead sample a grid of global intrinsic axis ratios as differences in angle do not always equally strongly translate in different intrinsic mass distributions. Following \citet{vanvan09}, we setup a grid in global intrinsic axis ratios ($p, q$), which we sample in steps of $0.1$ as well as in $u$, the ratio between observed and intrinsic size of the Gaussian. Here ($p, q$) denote the the intrinsic axis ratios of a triaxial ellipsoid with long, medium and short axis $a,b,c$: $p=b/a, q=c/a$.  Each value of ($p, q, u$) corresponds to a set of viewing angles ($\theta, \phi, \Psi$) under the assumption of a global ellipticity (which we assume to be 0.35), if a solution exist -- it is e.g. not possible that $q > q'$, with $q'$ the observed axis ratio of a Gaussian component. We note that these global values ($p, q$) are chosen as an efficient way of sampling the viewing angle space and do not necessarily reflect a single triaxiality for the whole galaxy. 

Since the photometry has a twist in the PA, there exist two physically different angle solutions for each $p, q, u$ pair with either positive or negative $\Psi$ \citep{vanvan09}. We therefore calculate for each  $p, q, u$  grid point two Schwarzschild models.

For the grid-based search, besides the three viewing angles, there are 3 other parameters that we include in our modelling; a central black hole with fixed mass M$_{\mathrm{BH}} = 2\times10^{9}$\Msun, a dark matter halo and a mass-to-light ratio M/L. We fixed the black hole mass while exploring the $\chi^2$ of the grid with different shapes. Experimenting with different black hole masses for diffferent shapes showed that the difference in $\chi^2$ between different grid points was mainly determined by difference in shape and only to a lesser extent by difference in black hole mass.   

For the inclusion of the dark matter halo in the dynamical model we tried two different strategies. The triaxial Schwarzschild code allows the inclusion of a Navarro-Frenk-White \citep[NFW;][]{NavFreWhi97} halo profile. This profile is parametrized in the Schwarzschild code by the ratio of M$_{vir}$/M$_{\star}$, with M$_{\star}$ the total stellar mass in the MGE, and a halo concentration $c$.
As the concentration of the profile for a galaxy like PGC 046832 should be low \citep[e.g.][]{WecBulPri02,ZhaJinMo03}, we fix the concentration parameter to $c = 3$. For the fraction of halo over stellar mass, we explore values of 1000, 1500 and 2000. We note that a model with DM fraction 1500 has a dark matter fraction of 45\% within a sphere of 10 kpc, and the enclosed dark matter mass surpasses the stellar mass beyond 18 kpc.For the I-band M/L ratio we explore 5 values between 2.88 and 4.16 \Msun/$L_{\odot}$. Although there are differences in the value of the best-fit $\chi^2$ between these models, these parameters do not make enough difference to radically change the location of the best-fit model.

\begin{figure*}
  \includegraphics[angle=0, width=.45\textwidth]{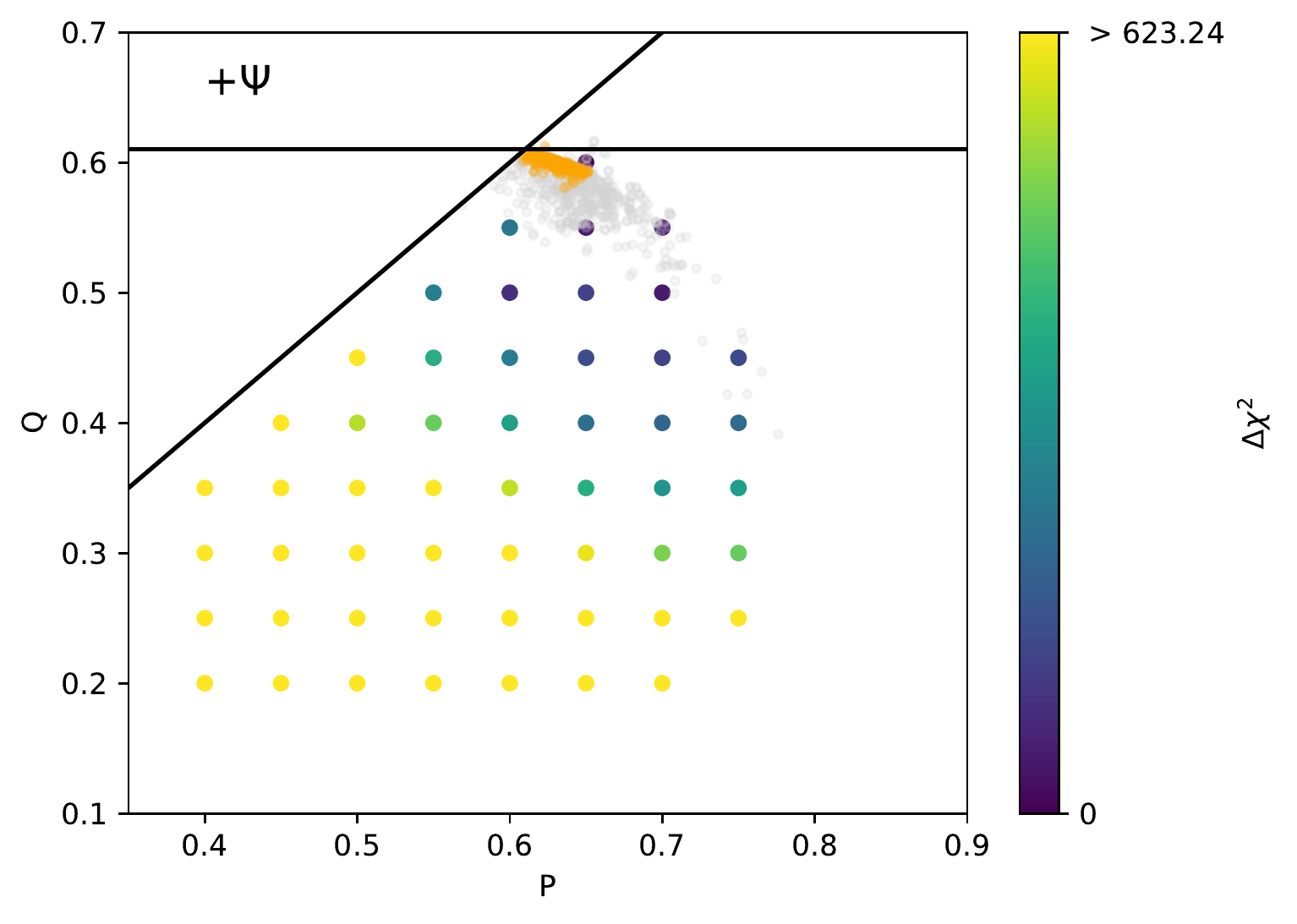}
  \includegraphics[angle=0, width=.45\textwidth]{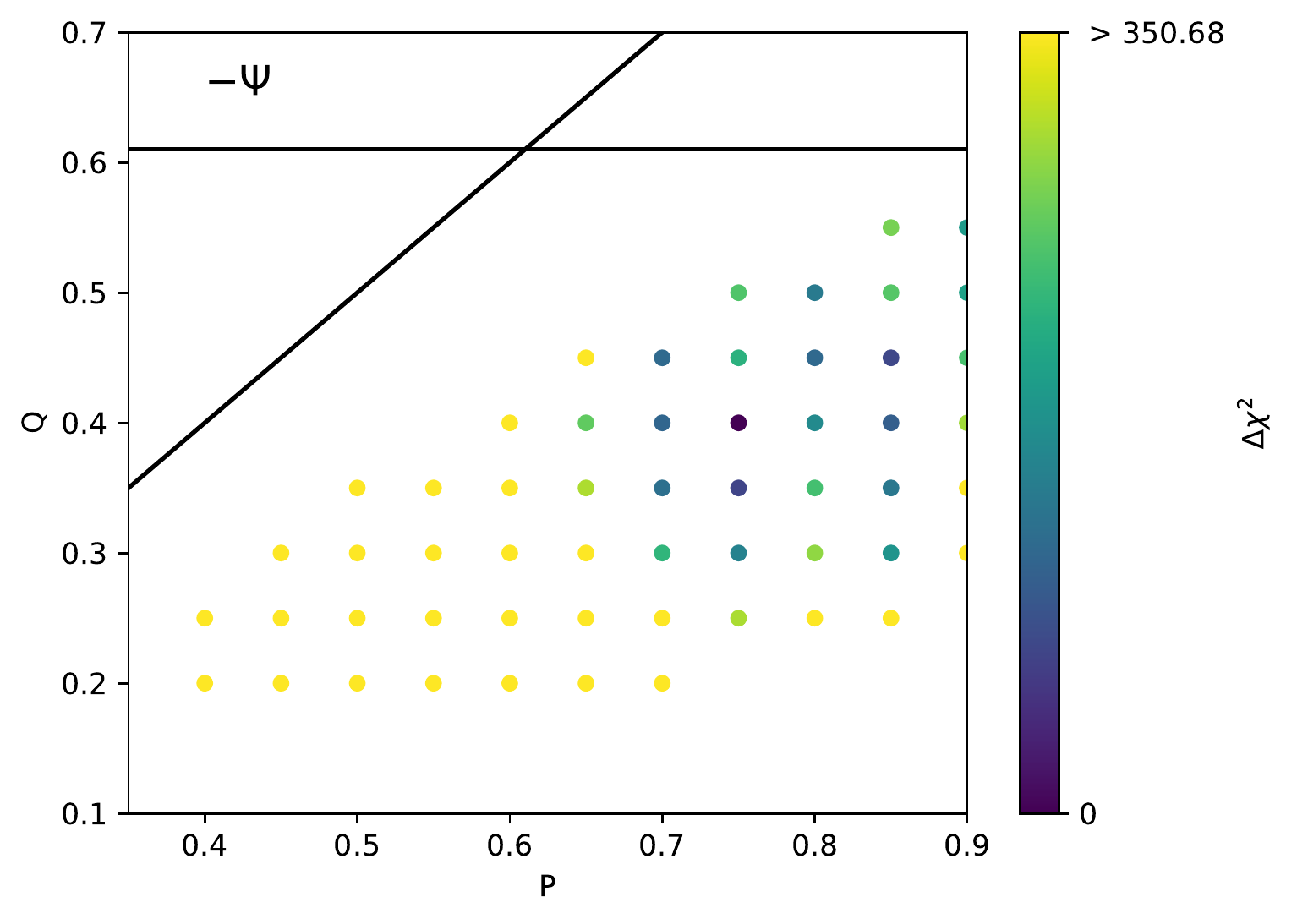} 
  \caption{$\chi^2$ distribution of models from sampling a grid of global values of $p$ and $q$, under the assumption of a global axis ratio $q' = 0.65$. The diagonal line denotes where models without triaxiality variation become prolate (note that the DM halo is spherical in all models), the
    horizontal line the maximally allowed intrinsic short axis ratio based on the photometry. Models are coloured by $\chi^2$ and marginalized over M/L and  DM fraction. As there are two different possible solutions for $\Psi$ because of the isophotal twist, we show the solutions with negative $\Psi$ on the right. The difference in $\chi^2$ between the best-fit models in the two plots is $250$. In the left panel we also show the distribution of galaxies from the ellipsoidal sampling search with grey dots. Models within 1-$\sigma$ from the best-fit model are marked orange.}\label{fig:shapechi}
\end{figure*}

Our second approach to find the best-fitting shape is based on  ellipsoidal sampling, a technique that is commonly used in nested sampling \citep{MukParLid06}. We start with a set of models (from the grid-based sampling), calculate the variance and covariance of the parameters (viewing angles, black hole mass and dark matter halo mass) of the 30 best-fitting models (excluding models which because of M/L ratio scaling have the same viewing angles and dark matter halos). We then find the covariance matrix of the parameters of these models, and determine the ellipsoid that encompasses the principal axes of this covariance matrix. We then predict using a multivariate Gaussian distribution (contrary to regular ellipsoidal sampling which uses a uniform distribution) the location of the next possible model. When a model is finished, the covariance matrix is updated and the process is repeated. This algorithm is not particularly fast at approaching the best-fit model, but explores the surrounding parameter space in a way that allows us to determine uncertainties on the viewing angles and dark matter halo parameter. Note that as we sample these models by viewing angle instead of ($p,q,u$), model parameters in Fig. \ref{fig:shapechi} can actually cross the $q'= 0.62$ line, since we assume a global ellipticity of 0.65. Also, instead of using a grid of M/L values, for each set of viewing angles, black hole mass and DM halo parameter, we try out four different values of the M/L ratio, and by fitting a second order polynomial to the three M/L values with the lowest $\chi^2$ value, predict a better fitting M/L value. We iteratively calculate the $\chi^2$ value for this predicted M/L, until the M/L is determined with an accuracy of 1 percent.

Fig. \ref{fig:shapechi} (left) shows the best $\chi^2$ at each position of the global $p, q$ (chosen to be the lowest $\chi^2$ among all models with those values of $p, q$), as well as for the solutions with negative $\Psi$ (right). The solutions with negative $\Psi$ have a much higher $\chi^2$ than the best-fit models in the left panel. We also show the models from the ellipsoidal sampling, where orange models are those within $1-\sigma$ from the best-fit model.

\begin{figure}
  \includegraphics[angle=0, width=.45\textwidth]{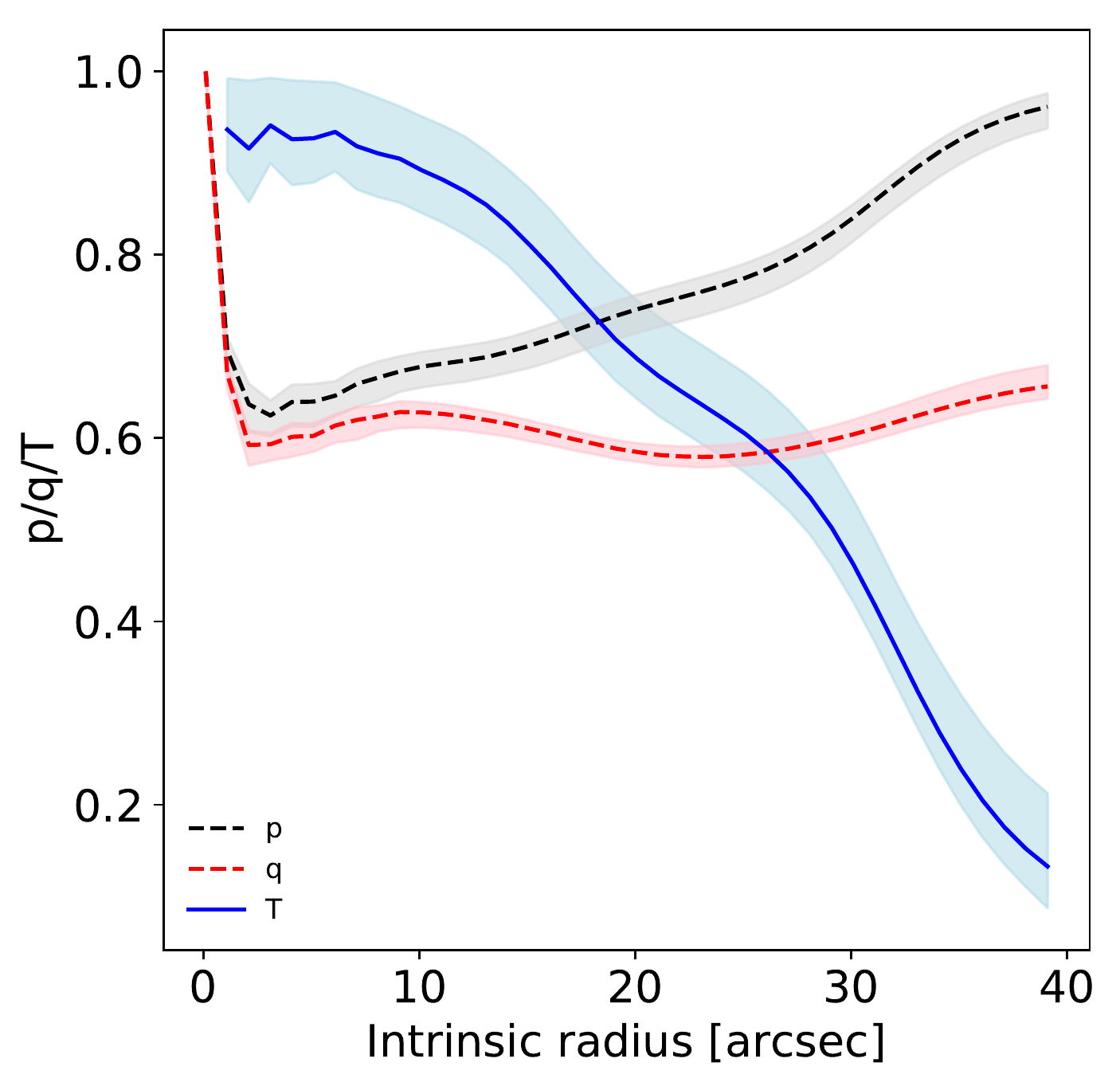} 
  \caption{Intrinsic shape of the model with the best-fit angles determined
    from kinematics. Intermediate axis ratio $p$, short axis ratio $q$ and
    triaxiality T 
  are shown as black dashed, red dashed and solid blue lines. The shape of the
  galaxy is close to prolate in the inner 5 arc seconds (T $\approx 0.9$) and
  becomes almost oblate in the outer parts.}\label{fig:intshape}
  \end{figure}

For completeness we note that the best fit viewing angles are ($\theta, \phi, \Psi$) = ($61.0_{-1.8}^{+3.8}$,$-59.4_{-2.4}^{+4.8}$,$74.9_{-1.2}^{+1.0}$). Using these angles we derive the intrinsic shape ($p_j, q_j, u_j$) of each MGE component. We note that the best fit values $u_j$, the ratio of the observed $\sigma$ over intrinsic $\sigma$ of each MGE component, are very close to 1 (between 0.94 and 0.99). In Fig. \ref{fig:intshape}, we show how the intrinsic axis ratios $p$ and $q$ change as a function of distance along the intrinsic major axis. The deprojected model of the stellar light is close to prolate in the inner $\sim 10$\arcsec, but becomes almost oblate in the outer parts (note that the dark halo is always spherically symmetric). This strong change in shape is the driver of the isophotal twist of $\sim10\degree$ seen in the photometry of the galaxy. We note that although the model is close to prolate at small radii, it is still able to support short axis tube orbits in the centre.

\begin{figure}
  \includegraphics[angle=0, width=.45\textwidth]{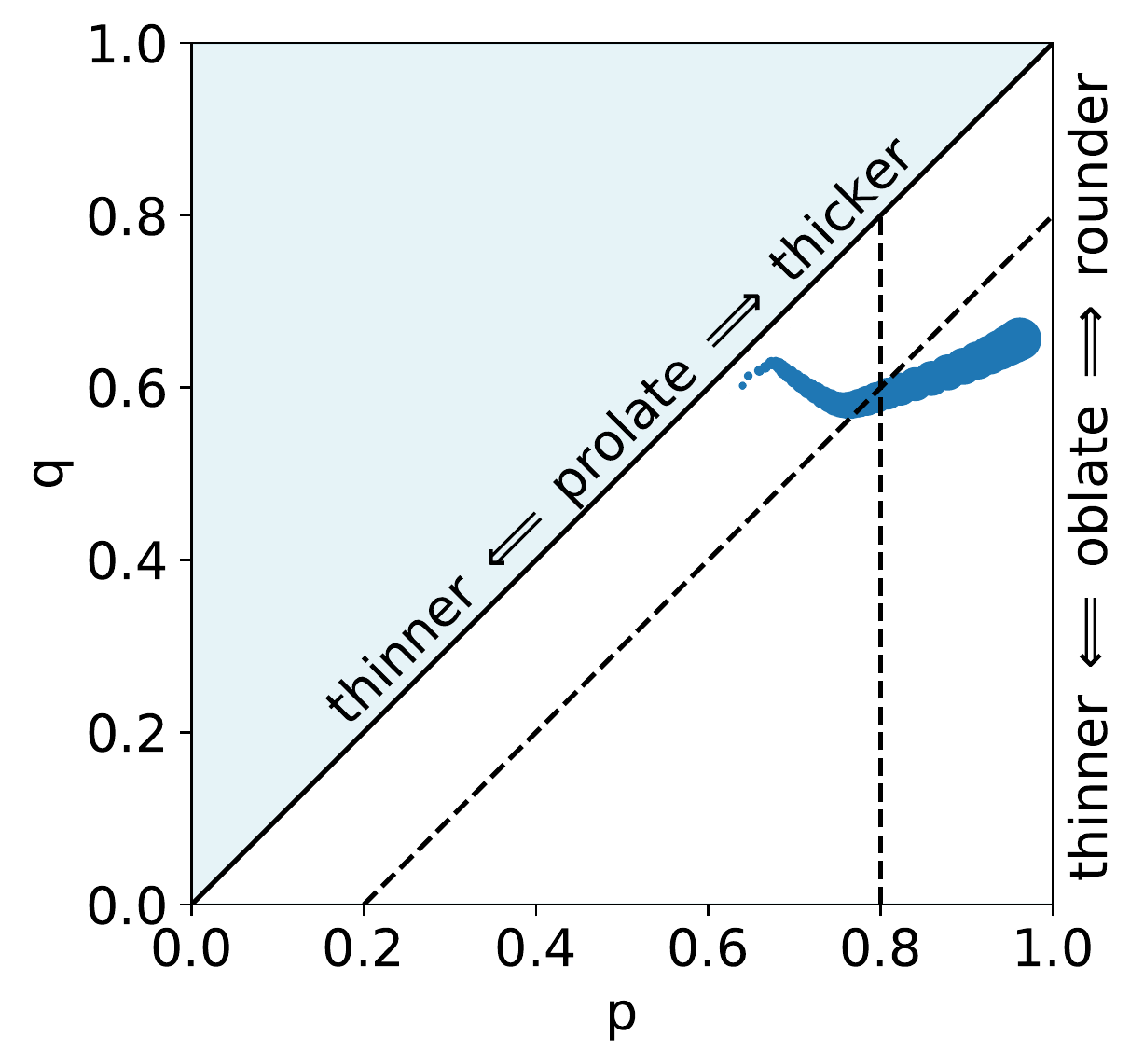} 
  \caption{Intrinsic short axis ratio $q$ vs intermediate axis ratio
    $p$ for the best fitting model. Bigger markersize denotes further distance
    from the galaxy centre. The light blue area is forbidden. With dashed lines we denote
    the area marked by \citet{LiMaoEms18} as prolate and oblate.}\label{fig:intshape_vsLi}
  \end{figure}

\begin{figure*}
  \includegraphics[angle=0, width=.98\textwidth]{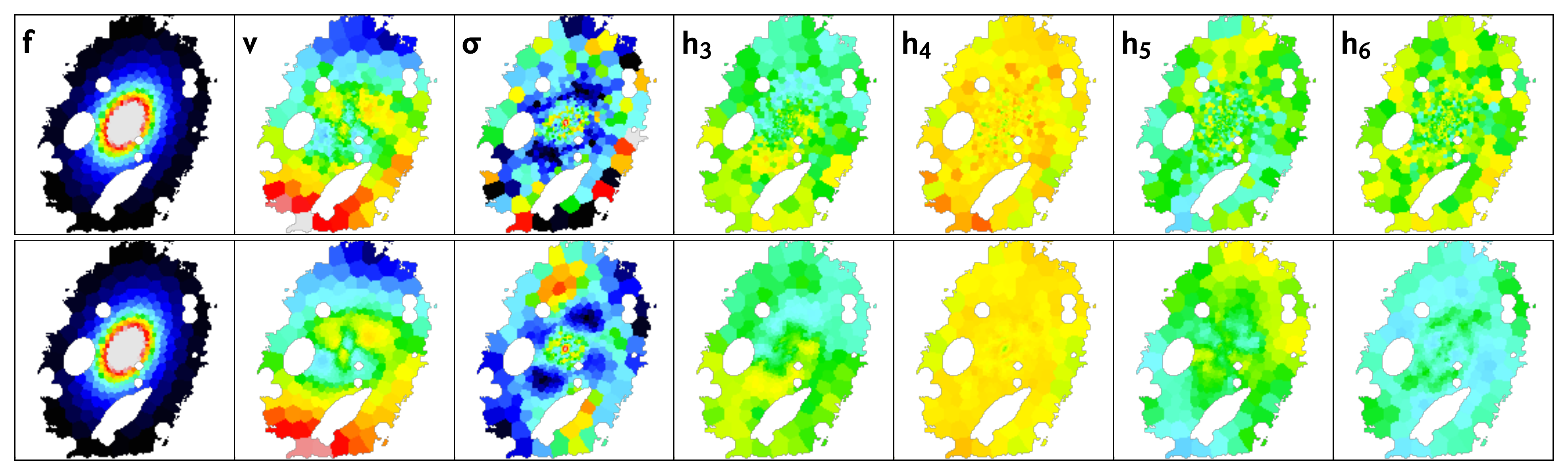}
  \includegraphics[angle=0, width=.98\textwidth]{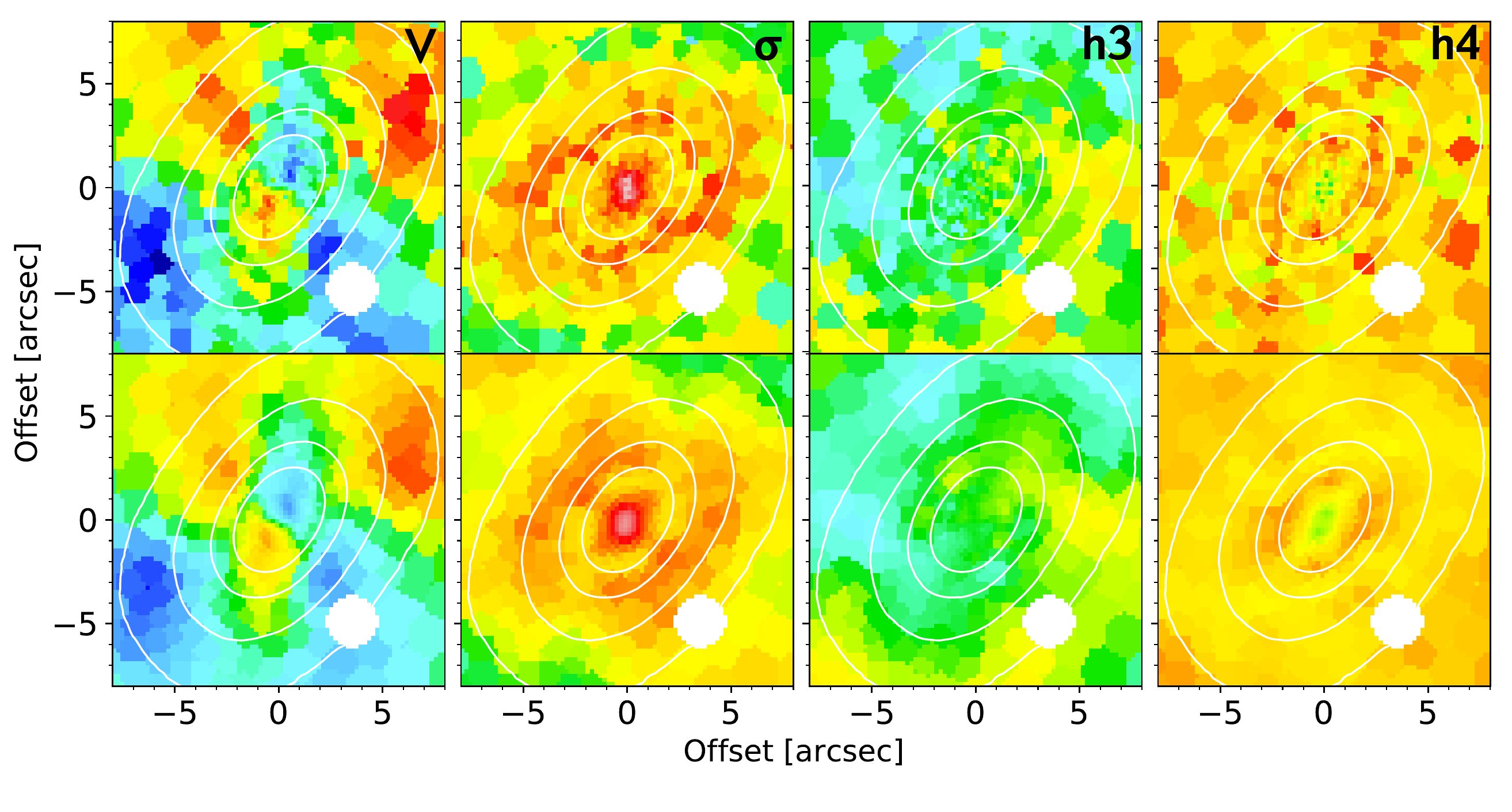}
  \caption{Observed stellar surface brightness and kinematics of PGC 046832 and the best fit kinematics of the Schwarzschild model. The first two rows show the full MUSE field-of-view, the lower rows the central AO corrected data. }\label{fig:shape_model}
  \end{figure*}

\subsection{Black hole mass}\label{sec:mbh}
If the kinematic reversals in PGC 046832 formed by the process seen in \citet{RanJohNaa19}, a high black hole mass is expected for this galaxy. We use three different codes to measure the mass of the black hole: the triaxial Schwarzschild code, an axisymmetric Schwarzschild code and axisymmetric Jeans models. We show here how different fitting codes lead to very different black hole masses.

\subsubsection{Triaxial Schwarzschild models}
For the BH mass measurement with the triaxial Schwarzschild code we assume the same parameters for the viewing angles and dark matter halo as for the determination of the shape, but we use the high-resolution kinematic data in the center and leave the black hole mass as a free parameter. We explore the grid of black hole masses and M/Ls shown in Fig. \ref{fig:mbh_contours}. The thick contours denote the models that are within $\sqrt{2N_{\mathrm{kin}}}$ from the best-fit model. We note that the  contours do not close on the low-mass end and therefore regard the best-fitting model as an upper limit. The sphere of influence of a $10^9$\Msun black hole is about 50 pc ($\approx$ 0\farcs05), well below the resolution of the data (seeing FWHM $\approx$ 0\farcs67).

\subsubsection{Jeans Anisotropic Models}
We determine the black hole mass also by fitting the second moment with Jeans Anisotropic Models \citep[JAM,][]{Cap08}.  These models predict the second velocity moment at each location in the galaxy, by solving the Jeans equations under the assumptions of 1) axial symmetry and 2) alignment of the stellar velocity ellipsoid to the cylindrical coordinate system. 
As input, the JAM models require an MGE for the surface mass density, an MGE for the the stellar luminosity density, an MGE of the PSF,  a value for the velocity anisotropy $\beta_z = 1 - \frac{\sigma_z^2}{\sigma_R^2}$ for each Gaussian component of the stellar luminosity density MGE (which we assume to be constant for the entire galaxy), a value for the inclination and the mass of the supermassive black hole.
We assume that the second velocity moment of PGC 046832 can be approximated by V$_{\mathrm{RMS}} = \sqrt{V_{\mathrm{LOS}}^2 + \sigma_{\mathrm{LOS}}^2}$, where $V_{\mathrm{LOS}}$ and $\sigma_{\mathrm{LOS}}$ are the velocity and dispersion re-fitted with \texttt{ppxf} to the observed spectra with higher order Gauss-Hermite coefficients fixed to zero. 
We fit JAM models with 5 free parameters to the V$_{\mathrm{RMS}}$ data. Besides the aforementioned black hole mass $M_{\mathrm{BH}}$, stellar anisotropy $\beta_z$ and inclination $i$, these also include a stellar mass-to-light ratio (M/L) and a NFW halo which we approximate by a $r^{-1}$ profile (and hence has only one free parameter). We explore this parameter space with the MCMC code {\it emcee}. A corner plot of the parameters can be found in Fig. \ref{fig:jam_mcmc}.
In Fig. \ref{fig:mbh_contours} we show the contours of the best-fitting models as a function of M/L and M$_{\mathrm{BH}}$. We find a best-fit mass for the central black of $\log$(M$_{\mathrm{BH}}$)$ = 9.96 \pm0.08$ (3-sigma uncertainty). The sphere-of-influence of a $10^{10}$\Msun black hole is (assuming $\sigma \approx 300$km/s) about 500 pc, which corresponds to roughly 0\farcs5 at the distance of the galaxy.

\subsubsection{Axisymmetric Schwarzschild models}
As a cross check, we determine the best-fit black hole mass using the axisymmetric Schwarzschild code of \citet{CapBacBur06,CapEmsBac07}. We use the same axisymmetric MGE as for the Jeans models and the same dark matter halo. The total mass of the dark matter halo was kept fixed to the best-fit mass of the triaxial modeling.  
The main aspects of the axisymmetric code are similar to the triaxial code. The model creates an orbit library of $21 \times 7 \times 8 \times 2 \times 6^3$ orbits, based on the mass model and assumed symmetry, where the last factor specifies the dithering scheme that improves on the smoothness of the distribution function. The orbital library is optimally superimposed and used to predict  V, $\sigma$ and Gauss-Hermite moments at the location (Voronoi bins) of the MUSE kinematics, which are then compared with the observations, providing a $\chi^2$ measure of the fit quality. We symmetrise the stellar kinematics by averaging the kinematic in the four quadrants defined by the major and minor axis of the galaxy ((x,y),(x,-y),(-x,y),(-x,-y)). When averaging we take into account that the odd moments of the LOSVD (V and h3) are bi-antisymmetric, while the even moments ($\sigma$ , h4) are bisymmetric. We retain the original kinematic errors at each point, not to underestimate the LOSVD parameter errors and overconstrain the model with artificially low formal errors. We use both the natural seeing and the AO assisted observations to constrain the Schwarzschild models, but we exclude from the fit the central 3\arcsec of the kinematics data set obtained under the natural seeing. 

\begin{figure}
  \includegraphics[angle=0, width=.45\textwidth]{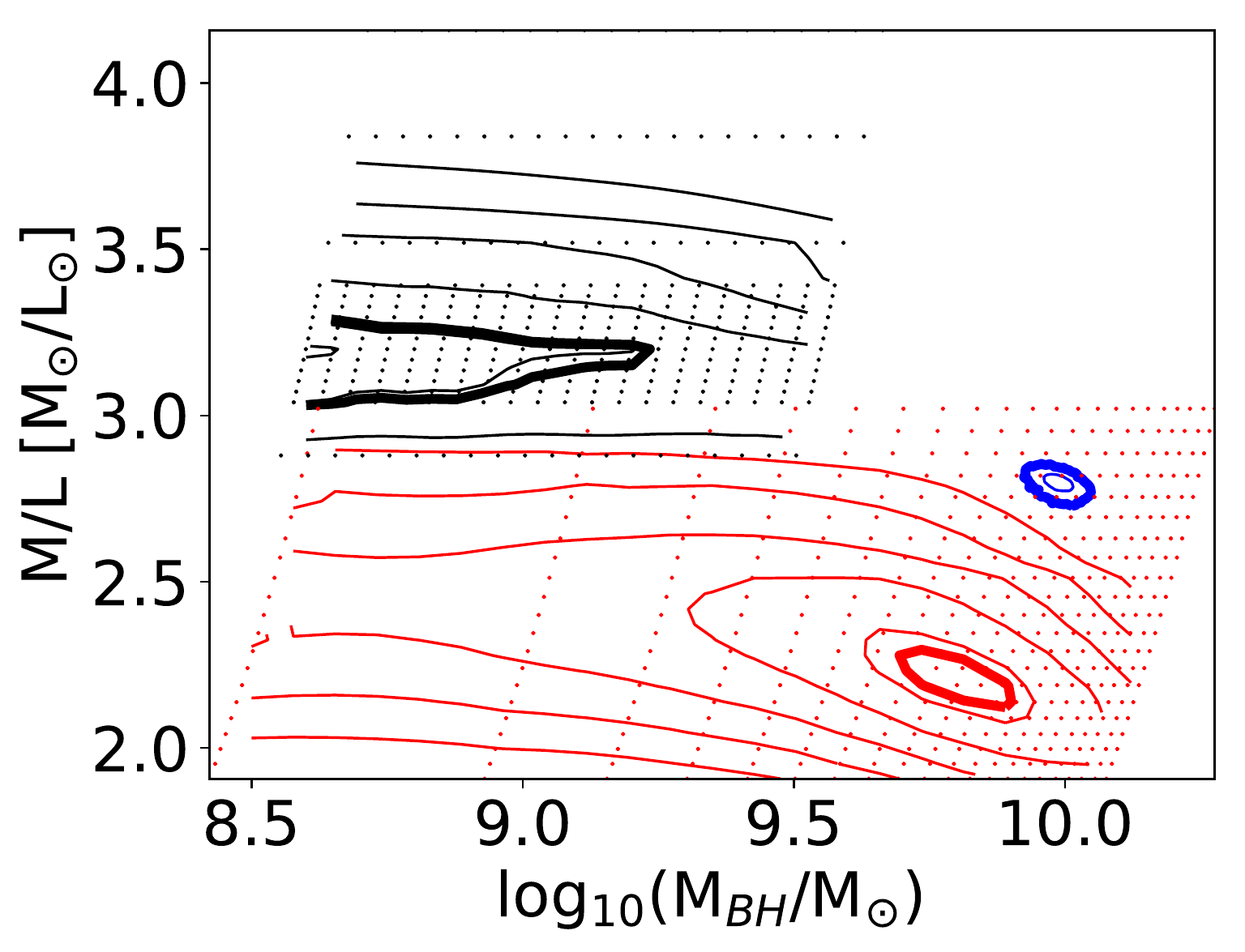} 
  \caption{$\chi^2$ map of mass-to-light ratio versus black hole mass derived
    from triaxial dynamical models (black), axisymmetric Schwarzschild (red) and Jeans models (blue) to the data. The thick black contour contains models within $\sqrt{2 N_{\mathrm{kin}}}$ from the best-fit model. The red contour contains the 3$\sigma$ best-fit models based on a $\Delta\chi^2$ criterion for 2 degrees of freedom. The blue contours are the (3$\sigma$ equivalent) 99\% confidence interval of the axisymmetric Jeans models. Red and black dots denote model grid positions of the Schwarzschild models.}\label{fig:mbh_contours}
\end{figure}

The mass model, next to the MGE inputs specifying both the distribution of the tracers (stars) and the distribution of the mass, is modified by an additional constant mass-to-light ratio and the mass of the black hole. These two parameters form a grid of Schwarzschild models used to search for the one with the smallest $\chi^2$ (Fig. \ref{fig:mbh_contours}).

The important differences between the axisymmetric and triaxial models are in the orbit library and in assumptions for the orientation of the galaxy. The axisymmetric orbit library allows for only short axis tubes, while there are no long axis tubes and box orbits. Even though there are orbits which have very low angular momentum (almost radial orbits) as well as orbits extending high above the equatorial plane of the galaxy, it cannot be expected from an axisymmetric model to reproduce the rotation around the major axis. The bi-(anti)symmetric data ensure that such prolate-like rotation is not present in the kinematics used to constrain the axisymmetric Schwarzschild models (as can be seen in Figs. \ref{fig:axi_schw_res} and \ref{fig:axi_schw_res2}). In the axisymmetric case, the orientation of the galaxy is fully specified with its inclination. Due to the inherent degeneracy of finding the inclination \citep{Ryb87, GerBin96, KraCapEms05}, we do not look for it, but assume the value of 85$\degree$.

The axisymmetric Schwarzschild models give as the best fit black hole mass log(M$_{\rm BH}$) = $9.76^{+0.14}_{-0.07}$ (three sigma level confidence limits). This mass is significantly lower than the mass recovered by the Jeans models, as well as much higher than the triaxial upper-limit M$_{\rm BH}$ estimate. The sphere-of-influence of a black hole with this mass would be $R_{SOI} \approx 0\farcs28$.

In Appendix \ref{apx:axi}, we show the symmetrised MUSE data, the best fit model and the residuals for both the AO and non-AO kinematics. The residuals are calculated as the difference between the model and the data, divided by the relevant uncertainties. As Figs. \ref{fig:axi_schw_res} and \ref{fig:axi_schw_res2} show, the model is in general able to reproduce the symmetrised observations, especially all the sign changes in the symmetrised velocity maps (both AO and non-AO data). In other kinematic maps, however, there are regions where the deviations approche a $3\sigma$ level. These are notably visible on the large scale map of the velocity dispersion and the $h_4$ Gauss-Hermite moment. The AO data are, generally, better reproduced, except again the $h_4$ moment map and some low-level residual structure in the velocity dispersion map. The central feature on the model velocity dispersion map (AO data) is interesting as it shows a double peak. This is a consequence of the counter-rotating structure (the double peak coincides with the innermost KDC), which the model reproduces by populating prograde and retrograde orbits, and it existence is not dependant on the black hole mass (i.e. it is also a fearure of models with low M$_{\rm BH}$). 

Data model comparisons of Figs. \ref{fig:axi_schw_res} and \ref{fig:axi_schw_res2} suggest that even under an axisymmetric assumptions (no twist in the photometry, no rotation around the major axis), the axisymmetric Schwarzschild model has difficulties reproducing all details of the symmetrized kinematics. Nevertheless, the mass of the black hole is robustly constrained, given the symmetrised data, as models with low $M_{\rm BH}$ (i.e. log(M$_{\rm BH}<$ 9.6) cannot reproduce the central velocity structure.

\section{Stellar population synthesis modeling}

\label{sec:ssp}
The stellar populations in PGC 046832 may hold clues to the formation of this galaxy. Our goal in this Section is to show how we derive both mass-weighted and light-weighted ages and metallicities as well as $\alpha$-element abundances [$\alpha$/Fe] and color profiles.

Although we note that it is possible to obtain values of [$\alpha$/Fe] using full-spectral fitting over an extended wavelength range \citep[e.g.][]{BarSpiArn20}, we instead opted for a classical approach by calculating absorption line indices. For each Voronoi bin we de-redshift the spectrum and convolve the spectrum to a resolution of 15\AA\ (FWHM). This resolution does not allow a comparison through the more common 14\AA\ MILES line index system \citep{VazSanFal10}, but the 14\AA\ resolution is slightly too narrow for the innermost bins wich have a dispersion over $\sigma \sim 340$ km/s. We measure the indices of H$\beta$, Fe5015, Fe5270, Fe5335 and Mg~$b$ using the definitions of \citet{TraWorFab98} and the methodology for calculating indices and variances outlined in \citet{TraFabDre08}.

Our choice of stellar library is the MILES library \citep{VazRicCen12}, which is sampled at a wide range of ages and metallicities (up to [Z/H] = 0.4), as well as at [$\alpha$/Fe]=0 and [$\alpha$/Fe]=0.4.  Using the same index definitions and resolution (15\AA) as for the data, we also measure indices on the MILES spectra. Using a linear interpolator, we predict for each combination of Age,[Z/H] and [$\alpha$/Fe] the values of the indices. We allow for extrapolation of metallicity and $\alpha$-enhancement up to [Z/H]=0.56 and [$\alpha$/Fe]=0.6. For each Voronoi bin, we then minimize a $\chi^2$ of all predicted indices.

\begin{figure*}
  \includegraphics[angle=0, width=.99\textwidth]{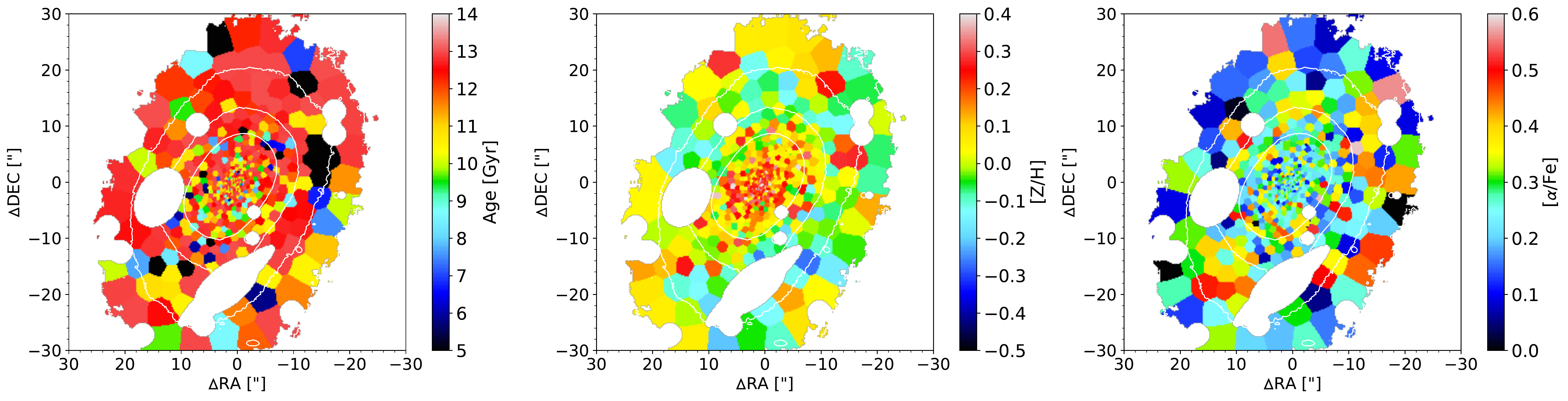}
  \caption{Maps of best-fit stellar population synthesis fitting results from index fits with MILES models. Panels show age, total metallicity ([Z/H]) and $\alpha$-enhancement [$\alpha$/Fe]. }\label{fig:ssp_par}
  \end{figure*}

We show the best-fit index-based ages, metallicities and [$\alpha$/Fe] values in Fig. \ref{fig:ssp_par} and as a one-dimensional plot in Fig. \ref{fig:ssp_1d}. We see a modest negative gradient in metallicity and almost no gradient in age or  [$\alpha$/Fe].

\begin{figure}
  \includegraphics[angle=0, width=.45\textwidth]{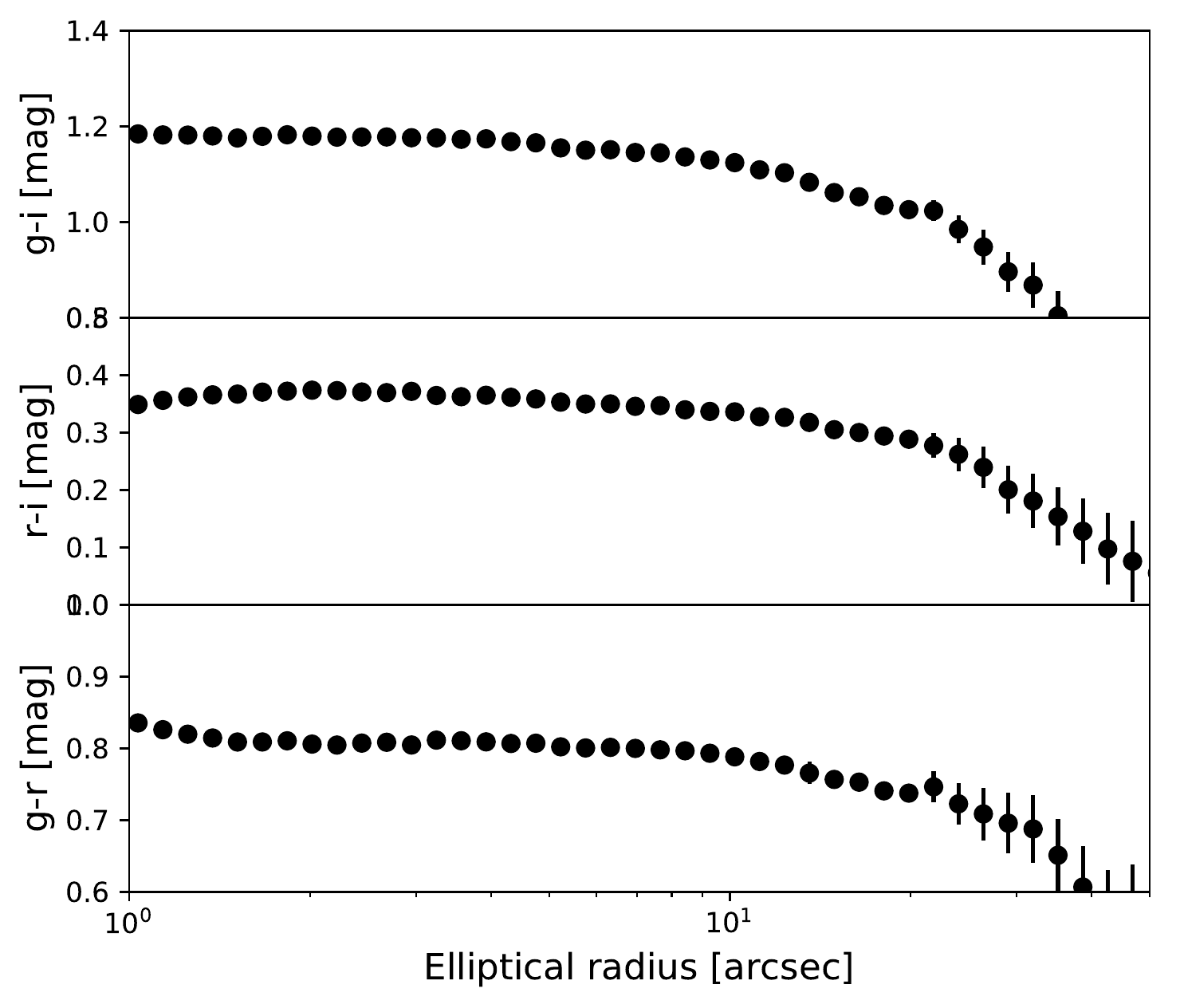} 
  \caption{Radial color profiles from the VST  {\it gri} band observations of PGC 046832. Colours are expressed in AB magnitudes. The profiles are steadily becoming bluer between 1\arcsec and 30\arcsec for all colours.}\label{fig:colorprofs}
\end{figure}

The VST/OmegaCAM data confirm that colour gradients in the inner parts of the galaxy are rather shallow.
We  measure radial intensity profiles of PGC 046832 from the VST images in the {\it gri} bands,
using \textsc{galphot} \citep{FraIllHec89}. We fix the ellipticity and
position angle to global values and then measure the surface brightness in
each band in the same elliptical apertures. We correct for small offsets in
position between the different bands by iteratively running \textsc{galphot}
and determining the best-fitting central position. Even though they are
similar in width, we do not attempt to homogenize the PSFs of the different 
bands, but instead  ignore the central $\sim 1$\arcsec. We apply a
k-correction to the colours to be able to compare them with M/L ratios
calculated for stellar populations at redshift $z=0$. The colour profiles are
shown in Fig. \ref{fig:colorprofs}, and show that the stellar populations are
becoming slightly bluer with radius, consistent with the shallow metallicity gradient from spectral fitting. All three colours show an even stronger radially
blue gradient beyond $\sim 20$\arcsec. Between $\sim 1$\arcsec -- $20$\arcsec, the {\it g-i} colour changes from 1.2 mag to 1.12 mag. This may be a manifestation of the outer S\'ersic component having different stellar populations.

\section{Discussion}
\label{sec:discussion}
\subsection{On the black hole mass discrepancies}
We have derived the mass of the black hole in PGC 046832 using
three different methods. The  oblate axisymmetric Jeans models
prefer a high ($\sim 10^{10}$\Msun) black hole mass. The axisymmetric
Schwarzschild models however favour black hole masses that are more in line
with the M-$\sigma$ relation (e.g. 4-6 $\times 10^9$ \Msun). The triaxial
Schwarzschild models provide only an upper limit to the black hole mass.

We first note that the assumption of a single galaxy-wide value for the anisotropy $\beta_z$ in the JAM models seems to have only limited influence on the black hole mass. Fixing $\beta_z$ to the values found from the best-fit triaxial Schwarzschild model, we find a black hole mass that is 0.05 dex lower for oblate JAM models. 

\begin{figure}
  \includegraphics[angle=0, width=.45\textwidth]{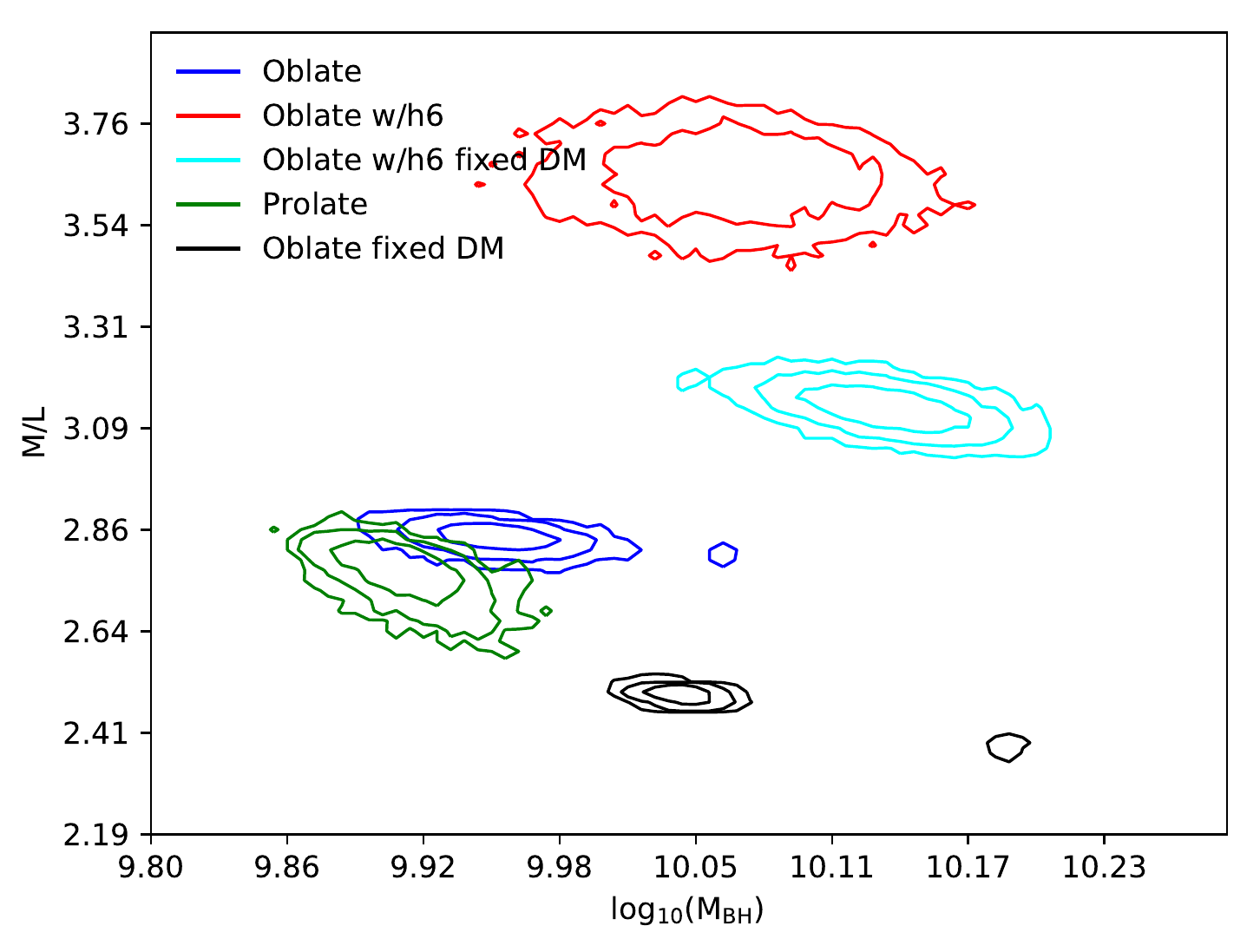}
  \caption{Contours of significance of the fit parameters for axisymmetric Jeans models. Contours contains 68, 95 and 99.7\% of the MCMC points. The blue contours at (M/L=2.86,M$_{\mathrm{BH}}=9.95$) show the likelihoods of oblate JAM models, the green contours at (M/L=2.7,M$_{\mathrm{BH}}=9.9$) show the likelihoods prolate models. Fixing the dark matter halo parameters to the values from the triaxial Schwarzschild model leads to a lower M/L and higher black hole mass as shown by the black contours. In cyan and red contours we show what happens when including higher-order Gauss-Hermite terms in the second moment. }\label{fig:mbh_contours_jam}
\end{figure}

The triaxial Schwarzschild models suggest that the intrinsic shape of PGC 046832 may be close to prolate in the inner parts. We therefore also fit prolate (axisymmetric) Jeans models to the data. This requires minor modification to the input of the JAM models (e.g. $\sigma \rightarrow q' \sigma$, $q' \rightarrow 1/q'$, and a rotation of 90 degrees of the orientation of the input data.) The results of the fits of these prolate Jeans models are also shown in Fig. \ref{fig:mbh_contours_jam}, and show only a minor difference for the best fit black hole mass.

Another difference between the triaxial models and the Jeans models is the normalization of the dark matter halo. If we fix the dark matter halo normalization to the halo mass found with the triaxial models instead of leaving the normalisation of the dark matter halo free, the JAM models prefer an even higher black hole mass and lower M/L, as shown by the black contours in Fig. \ref{fig:mbh_contours_jam}. This lower M/L is however consistent with the M/L of the axisymmetric Schwarzschild models, for which we had fixed the mass of the DM halo to the one in the triaxial models. \citet{ThoJesNaa07} have shown that the M/L of triaxial systems derived through axisymmetric methods can be biased low. This may provide an explanation for the lower M/L in the axisymmetric models and the lower halo mass for the Jeans models.

The Jeans models make use of the approximation $V_{\mathrm{RMS}}^2 = V^2 + \sigma^2$ for the second moment. The data to constrain the M/L and black hole mass with Jeans models are thus different from the data to which the Schwarzschild models were fitted. As a case in point we note that in the kinematic maps in Fig. \ref{fig:datakin_full_sub},  $h_4$ is seen to be slightly lower at the centre ($h_4 \approx 0$) than in the surrounding Voronoi bins ($h_4 \approx 0.05$). We  recalculate the second moment according to the formulas in Appendix \ref{app:gh}, and refit these second moments with JAM models. As can be seen from Fig. \ref{fig:mbh_contours_jam}, the inclusion of the Gauss-Hermite moments leads in fact to higher M/L and black hole mass. 

In general BH mass measurements with Jeans and Schwarzschild models for systems with a well-resolved sphere of influence do
agree \citep[e.g.][]{SetvanMie14,ThaKraCap19}. The result in this paper
resembles the result of \citet[][their Fig. 10]{AhnSetCap18}, where a
significant M${_\mathrm{BH}}$ detection with Jeans and axisymmetric Schwarzschild models disappears when using triaxial Schwarzschild models. The potential for the models assumed by Ahn et al. is close to axisymmetric, but still allows box orbits and long axis tubes. A similar discrepancy was seen by \citet{LieQueMa20}, where the removal of a slight triaxiality in the models led to a significantly higher black hole mass, which was then found to be consistent with axisymmetric Jeans models. It is clear that in the case of PGC 046832 there is no good argument to enforce axisymmetry as PGC 046832 is clearly triaxial, and both box and long axis tube orbits can be present.

\begin{figure}
  \includegraphics[angle=0, width=.48\textwidth]{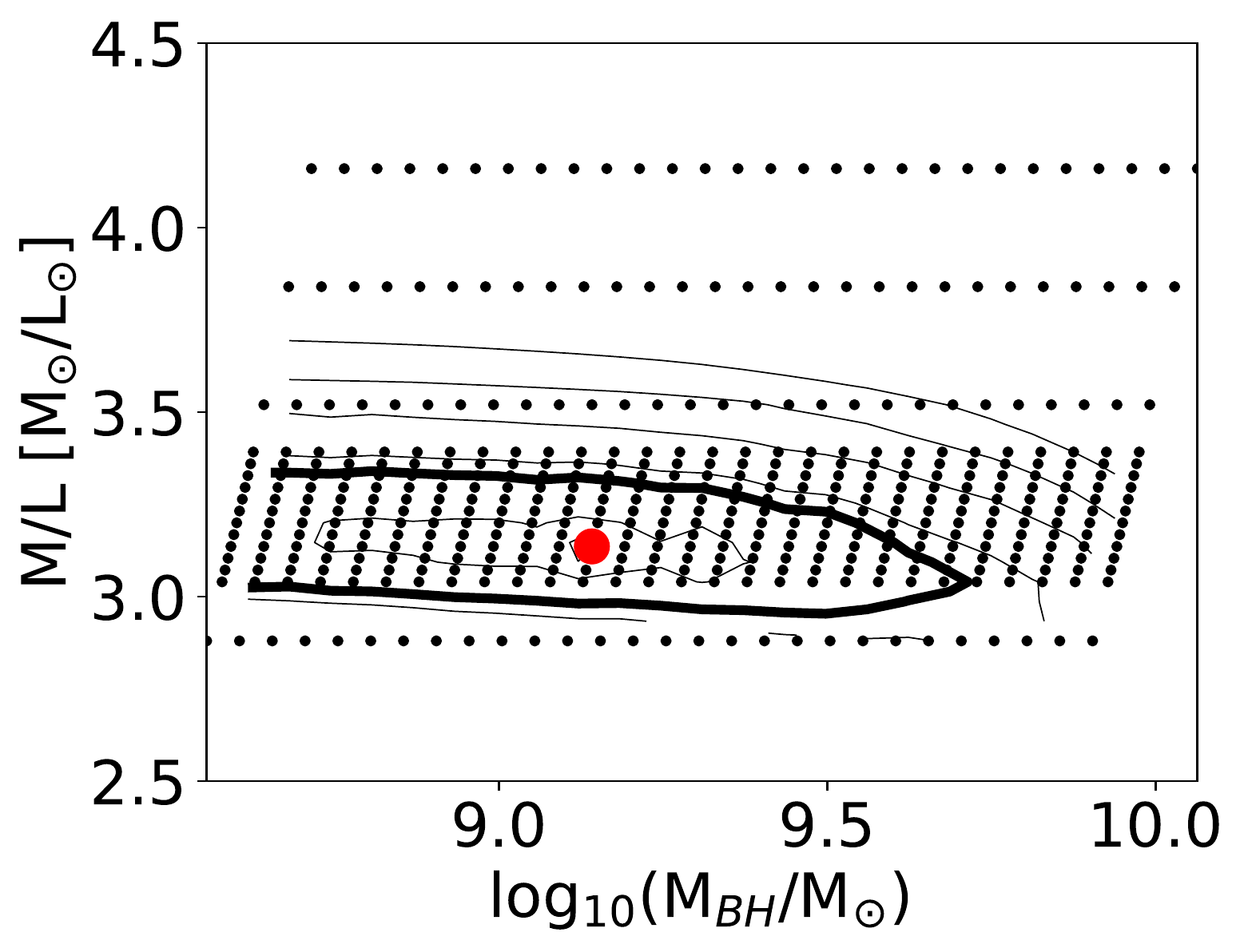}
  \caption{Results from triaxial Schwarzschild models where low-energy box orbits were excluded. This exclusion allows a detection of the black
    hole mass. Contours are as in Fig. \ref{fig:mbh_contours} }\label{fig:comp_nobox}
\end{figure}

In the case of PGC 046832, the sphere of influence is only marginally resolved even with Jeans models. This allows the triaxial Schwarzschild codes to fit the kinematics in the centre  with orbits that provide a high dispersion but do not necessarily require a high black hole mass. Taking away these orbits naturally leads to higher black hole masses. As an example, we show in Fig. \ref{fig:comp_nobox} $\chi^2$ contours for triaxial Schwarzschild models  for which we restricted the box orbits to have a higher orbital energy (which is equivalent
to radius), to avoid these orbits from starting too close to the centre. For this we extended the minimum starting radius of  0\farcs03 to 1\arcsec, while keeping the same parameters for the tube library\footnote{As the starting positions of orbits are derived by approximating the  potential as a separable potential, it is possible that some of the orbits in the tube orbit library are in fact box orbits and vice versa.}. In this case, box orbits close to the black hole are constrained through their photometry and kinematics outside the central resolution elements. As the figure shows, this allows the detection of a supermassive black hole.
The solution to the black hole mass discrepancy may therefore be acquiring higher-resolution kinematic data. In Appendix \ref{sec:cat} we show however that even with a resolution of 0\farcs5 the discrepancy remains.

\begin{figure*}
  \includegraphics[angle=0, width=.45\textwidth]{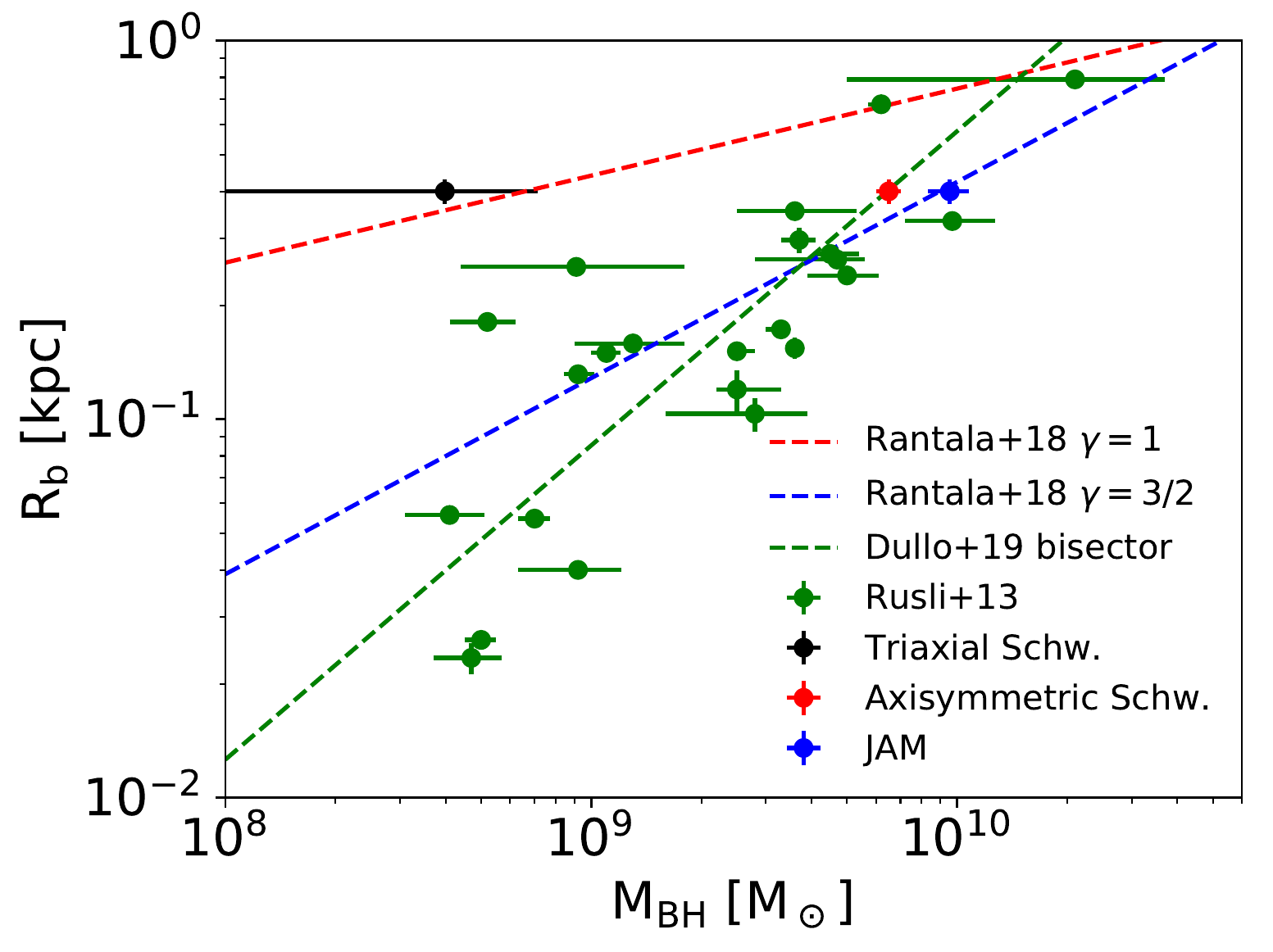}
  \includegraphics[angle=0, width=.45\textwidth]{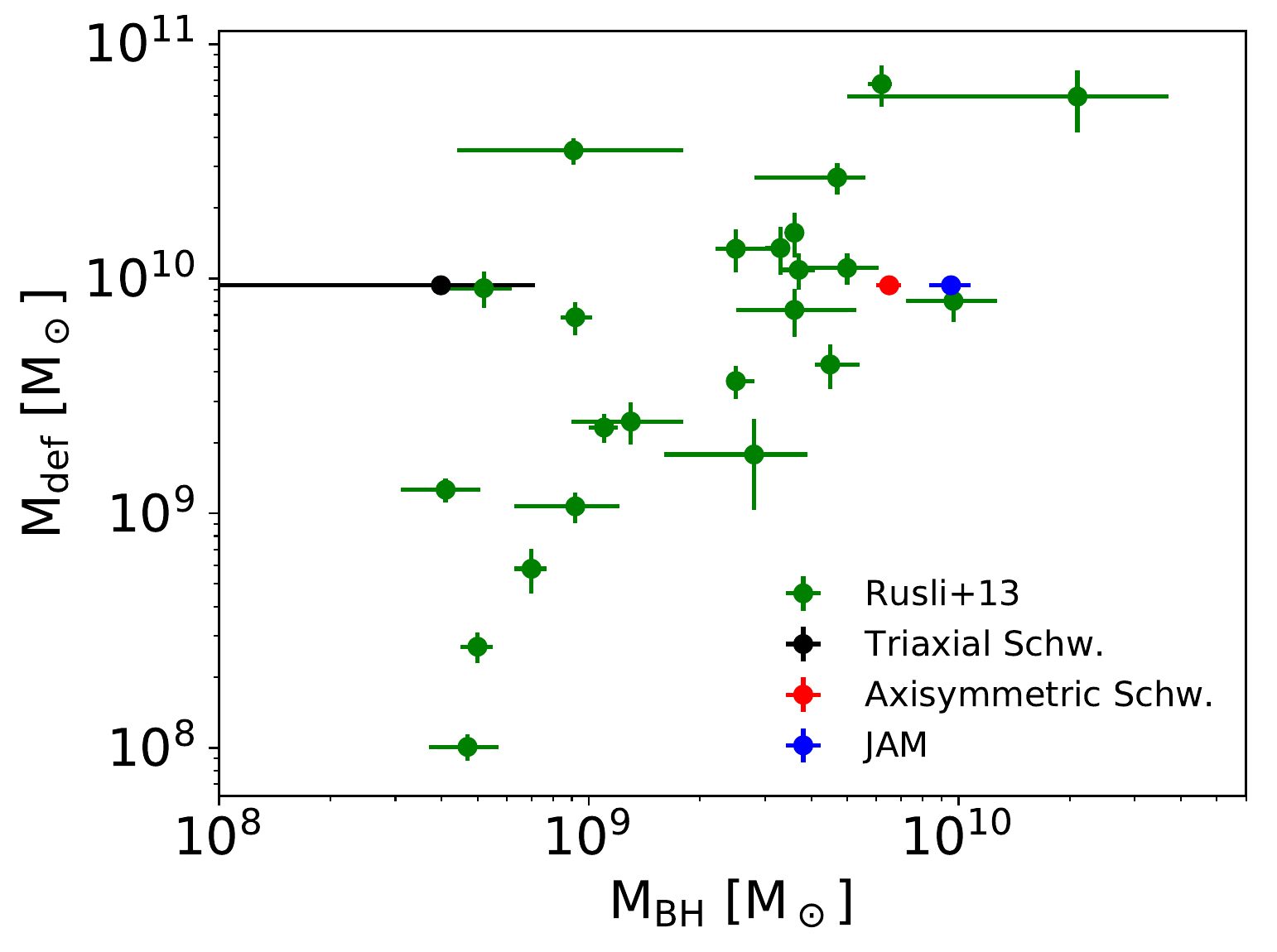} 
  \caption{Comparison of the inferred black hole masses of PGC 046832 with known scaling relations. Left we show the core radius versus black hole mass, right the mass deficit in the core versus the black hole mass. The green literature points were taken from \citet{RusErwSag13}. The locations of the triaxial and axisymmetric Schwarzschild and Jeans models are shown in black, red and blue (from left to right). Models of \citet{RanJohNaa18} are shown in blue and red dashed lines. \citet{Dul19}'s bisector fit is shown as a green dashed line. }\label{fig:mbh_rusli}
\end{figure*}

In Fig. \ref{fig:mbh_rusli} we show the inferred black hole masses from different methods versus the core radius, together with a literature sample of \citet{RusErwSag13} and scaling relations from the N-body simulations of \citet{RanJohNaa18}, for different initial stellar density profiles ($\gamma$ = 1, 3/2). For the size of the core of PGC 046832, the triaxial upper limit on the black hole mass may at first sight seem low. The models of \citet{RanJohNaa18} show that with progenitor galaxies with shallow density profiles ($\gamma = 1$), even a black hole binary with a final mass a little bit below 10$^9$\Msun is able to scour such a core. It is unclear if such shallow density profiles are expected for the progenitor galaxies of BCGs. Another explanation for the low black hole mass may be gravitational recoil from a massive black hole merger event. By such an event, the magnitude of which depends on the masses and spin alignment \citep[e.g.][]{CamLouZlo07}, the black hole may be ejected from the centre of the galaxy. In a triaxial galaxy, a massive black hole will spiral in rather quickly after such ejection, but may oscillate for extended time in the galaxy core \citep{GuaMer08}. This mechanism has been used to explain the unusually large core of the BCG galaxy in Abell 2261 \citep{NasGuaRea21}. A third explanation may be that the final black hole formed through a sequence of multiple mergers. \citet{Mer06} estimates that the missing mass is related to the black hole mass through $M_{\mathrm{BH}} \approx 0.5 \mathcal{N} M_{\mathrm{def}}$. Using the best-fit M/L ratio of the triaxial Schwarzschild model and the light deficit from the core S\'ersic model, we find a scoured mass $M_{\mathrm{def}} = (9.37 \pm 0.17) \times 10^{10}$\Msun. A plausible explanation may thus be that the discrepancy between core size and black hole mass is caused by $\sim 5$ mergers with black hole masses $M_{\mathrm{BH}} \sim 2\times10^8$\Msun.

\begin{figure}
  \includegraphics[angle=0, width=.49\textwidth]{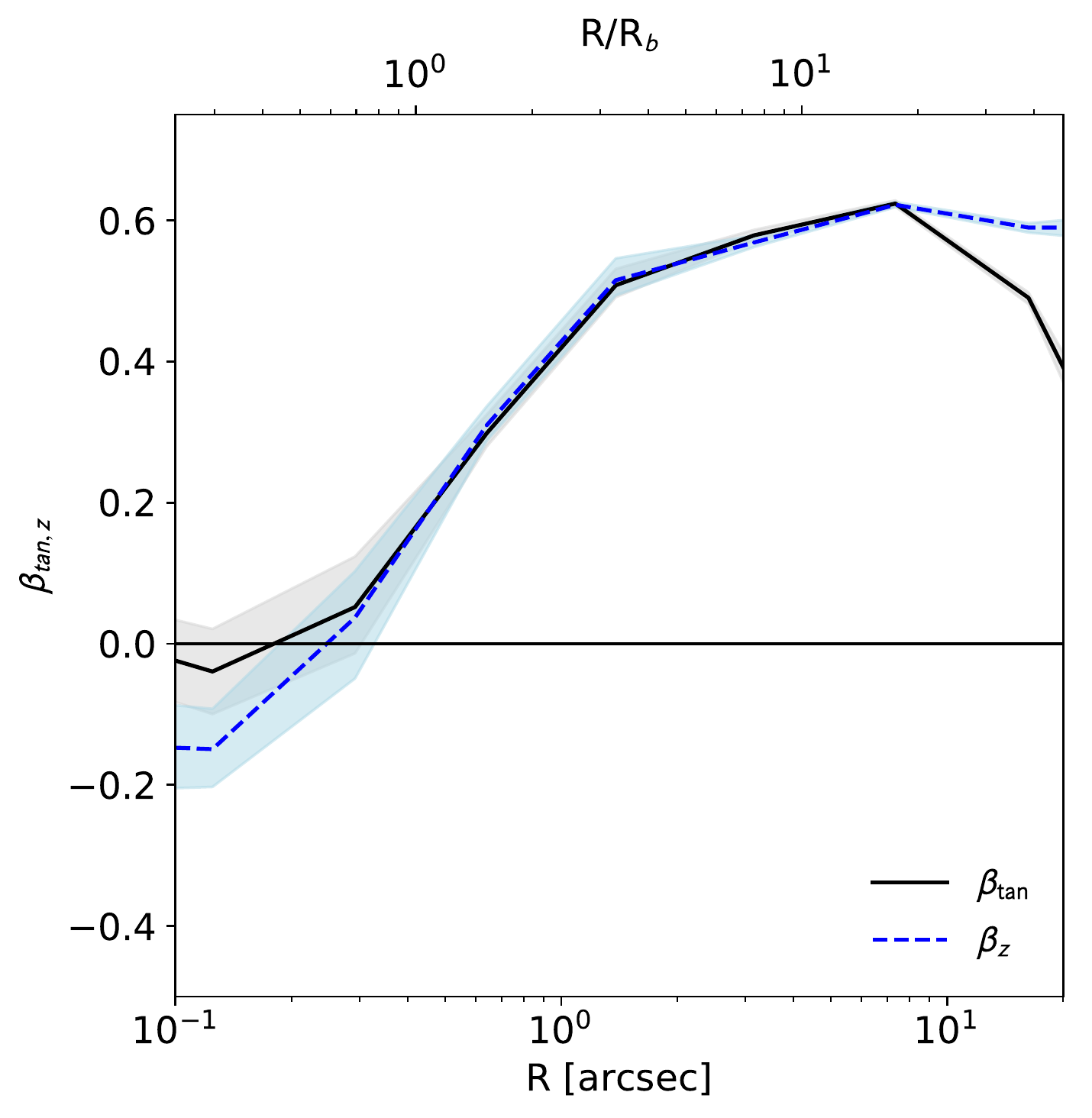} 
  \caption{Average orbital anisotropy in the triaxial Schwarzschild models  that are consistent with the best fit model as a function of radius. The black line shows the tangential anisotropy, the blue dashed line show $\beta_z$. The confidence intervals show the observed range of anisotropies in the models.}\label{fig:orbit_beta}
\end{figure}

In Fig. \ref{fig:orbit_beta} we show the orbital anisotropy $\beta_{\mathrm{tan}} = 1 - \frac{1}{2} \left(\sigma_{\phi}^2 + \sigma_{\theta}^2\right)/{\sigma_r^2}$ and $\beta_z = 1 - \sigma_{z}^2/\sigma_R^2$, as measured along the intrinsic major axis of all models consistent with the best-fit model of run 2.  Although the spatial resolution of our kinematic data is insufficient to probe the core in detail, the Schwarzschild models do show a tangential bias inside the core, expected from black hole binary scouring  \citep[e.g.][]{MilMer01} but significantly weaker than what has been observed with axisymmetric models \citep[e.g.][]{ThoSagBen14}. A weak tangential anisotropy can also be the result of repeated merging \citep{RanJohNaa19}.  We note however that this signal cannot be exclusively attributed to black hole scouring as other processes, such as star formation can also produce the tangential anisotropy.

\subsection{Shape}
We find a shape that is close to prolate in the inner 10\arcsec of the galaxy and becomes close to oblate at larger radii. This shape is based on comparing the kinematics of the Schwarzschild models with the observed kinematics, for different viewing angles, dark matter halos etc. To check that our methodology is robust for finding the best-fit shape, we perform two additional test. 1) We run 250 Schwarzschild models with an increased number of orbit bundles ($n_E \times n_{\theta} \times n_{R} =n_E \times n_{\theta} \times n_{phi} = 31\times15\times15$) using the ellipsoidal sampling method around the best-fit models. This did not lead to a change in best-fit viewing angles. 2) For a small subset of 21 models with the grid of M/L as in run 1, but a fixed dark matter halo ($M_{\mathrm{vir}}$/$M_{\mathrm{MGE}}$ = 1500), and fixed black hole mass ($M_{\mathrm{BH}} = 2 \times 10^9$\Msun) as in the grid-based search, we determine the effective number of free parameters using the methodology outlined in \citet{LipTho21}. For this we use $K=20$ realizations of each single model, and the Akaike Information Criterion approach, in which two times the number of effective parameters is added to the $\chi^2$. Calculating a new $\chi^2$ for the best fit model from the grid search shows that in order to find a model that might fit better, we can limit this calculation to a subset of 21 models $m$ that have $\chi^2_m < \chi^2_{\mathrm{best}} + 2 m_{\mathrm{eff,best}}$.
  Although this makes the $\chi^2$-distribution in $(p,q)$ shallower, it does not change the location of the best-fit model from the grid search, as can be seen in Fig \ref{fig:chi2_lipka}.    
One critical assumption that we made is that the observed light profile can be deprojected using the MGE formalism. This formalism assumes that the intrinsic density of the galaxy can be written as the sum of concentric, ellipsoidal and aligned Gaussian density distributions. The MGE formalism provides an intrinsically smooth, and quickly-calculated density distribution. However, the viewing angles of a galaxy as a whole are limited to those viewing angles that are consistent with each individual MGE component. Recently, \citet{DeNSagTho20} have presented a non-parametric density estimator based on nearly-ellipsoidal shells and compared its performance with MGEs, finding that intrinsic densities are better recovered with their non-parametric method than with the MGE formalism. It is possible that, despite the precautions we have taken in allowing as much viewing angles as possible, the viewing angles are biased or that the density distribution is sub-optimally recovered; however, the non-parametric methodology of  \citet{DeNSagTho20} is at this point not compatible with the Schwarzschild code.

A second assumption is that the galaxy is a single triaxial system. It has been observed in simulations that stars in the outer parts (beyond $\sim 1$ effective radius) trace the shape of the dark matter halo \citep{PulGerArn20}, of which the axes are not necessarily aligned with those of its central galaxy.

The main change of shape, from prolate to oblate, is driven by the strong isophotal twist observed in the photometry.
\begin{figure}
  \includegraphics[angle=0, width=.48\textwidth]{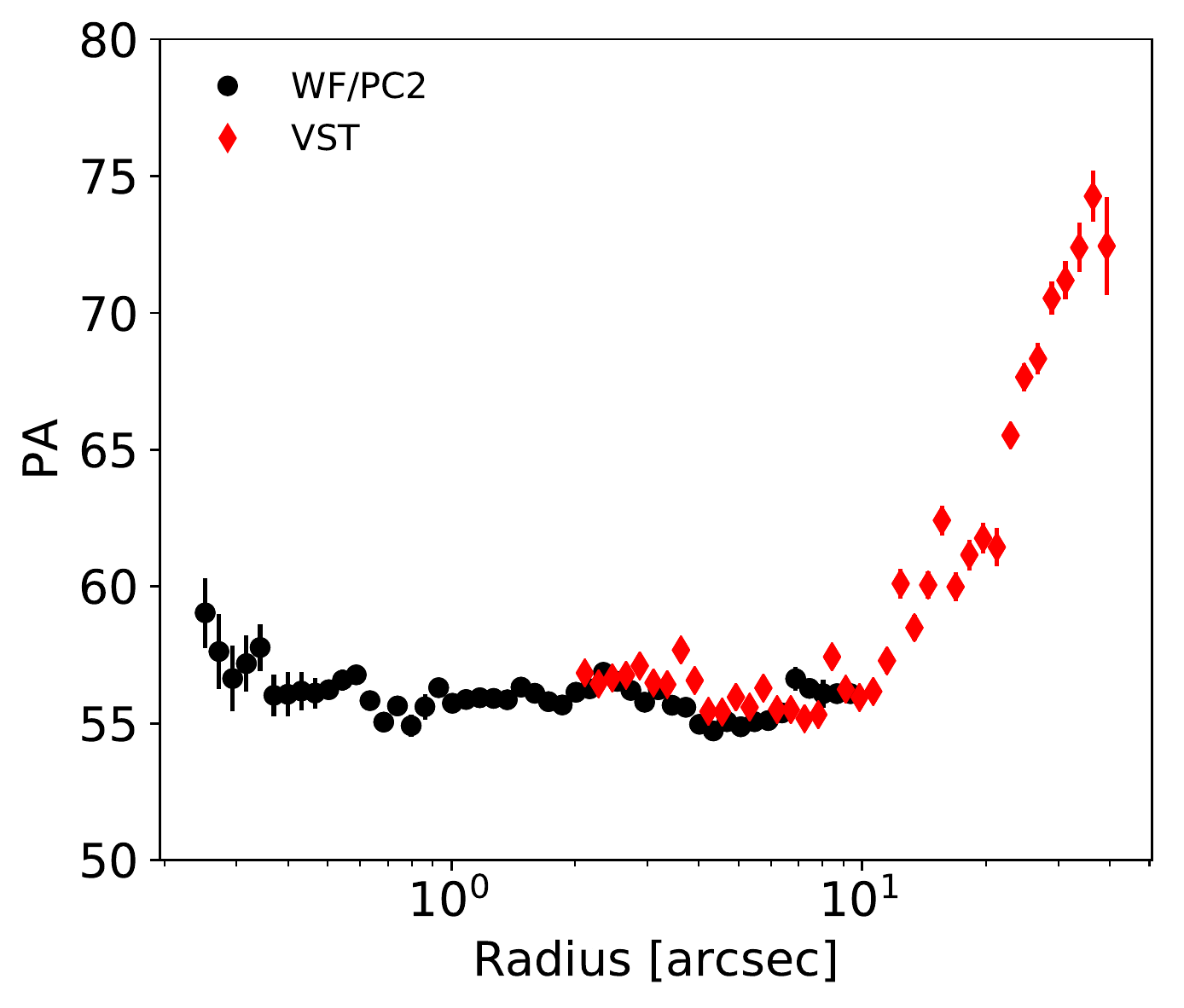}
  \caption{The P.A. of PGC 046832 as a function of radius. Black circles are measurements from HST WFPC2 data, red rhombi are the measurements from the VST data. }\label{fig:PA}
  \end{figure}
Fig. \ref{fig:PA} shows that the change in position angle, which we measured with \textsc{galphot} \citep{FraIllHec89} on the HST and VST photometry data, 
 happens abruptly at $\sim10$\arcsec. This is about twice the effective  radius of the central core-S\'ersic component and coincides with the region where the outer S\'ersic profile starts to dominate the light.

 A second S\'ersic profile for massive ellipctical galaxies has sometimes been interpreted as accreted mass \citep[e.g.][]{SpaIodvan20} but also as intracluster light \citep[e.g.][]{KluNeuRif20}. As the velocity dispersion of this component is lower than that of the cluster dispersion, the rotation seems regular, and we do not see evidence for asymmetry in position or structure of the outer parts,  we prefer the first interpretation. In case the stars in the second S\'ersic component, which shows the isophotal twist, are not in equilibrium with the main galaxy, we would not be able to use the isophotal twist to limit the viewing angles necessary for the deprojection of the MGE. Recent work on galaxies in the MASSIVE survey by \citet{GouJenBla18} shows that galaxies with isophotal twists up to 20$^{\circ}$ are not uncommon among massive galaxies. We can however not distinguish if in this case the twist is caused by a non-relaxed structure or if it is due to the triaxiality of the galaxy.  

 Prolate shapes are almost exclusively made in major dry mergers \citep[][]{LiMaoEms18}, as mergers with gas lead to oblate shapes \citep[e.g.][]{HofCoxDut10}. One way to create the observed shape would be the combination of a major dry merger, followed by the accretion of satellite galaxies, which simulations show would mainly be deposited in the outer parts \citep[e.g.][]{NaaJohOst07}. The mass-metallicity relation and lower metallicity in the outer parts imply that these satellites must have been lower mass than the progenitor galaxies forming the central component. However, the low S\'ersic index and lower metallicity of this second component may also arise from in situ formation. 

\citet{PulGerArn20a} present shape measurements for the outer halos (beyond one effective radius) of massive galaxies in simulations. Their analysis of high-mass slow-rotators in simulations shows that these galaxies are almost always prolate in their outer parts (8 effective radii). It is possible that PGC 046832's halo is somewhat unusual, but we note that perhaps the data are simply not reaching far enough out as they do not exctend beyond two effective radii.

\subsection{Orbit distribution and origin of the kinematically decoupled components}
In triaxial galaxies different types of orbits are found: long axis tubes, short axis tubes and box(let) orbits \citep{Sch93}.  42 per cent of the orbital weights in the best-fit triaxial Schwarzschild model (in the best fit model of run 2, using the ellipsoidal parameter optimization), which we consider our best-fit model, are in short axis tube orbits.
34 per cent of the orbital weights is given to long axis tube orbits. The remaining  23 per cent of the orbital weights is given to box or radial orbits. One concern for the calculation of the amount of box orbits is that, even though we only use the $\chi^2$ based on kinematics to find the best-fit models, the orbits are optimized to simultaneously reproduce kinematics {\it and} the projected and deprojected light distribution. As the latter is based on the MGE, of which the components have a perfectly elliptical shape, one might expect that this method discourages the non-linear least squares code to use box orbits. We find however that directly using the observed photometry does not change the amount of weight in box orbits or the best fit shape. \citet{JesNaaBur05} quantify the promince of box and tube orbits in merger remnants. Although a direct comparison between our model and their model is difficult because of the radially changing shape, they find for triaxialities above $T \gtrapprox 0.2$ that the amount of box orbits is between 20 and 40  per cent. However, \citet{RotNaaOse14}, find much higher fractions of box orbits for intermediate triaxialities. Although it is unclear how realistic the orbit distribution of the triaxial Schwarzschild model is, a comparison of orbit types between Schwarzschild models and simulation by \citet{XuZhuGra19} shows that in general Schwarzschild models reproduce simulated galaxies well.

\begin{figure}
  \includegraphics[angle=0, width=.49\textwidth]{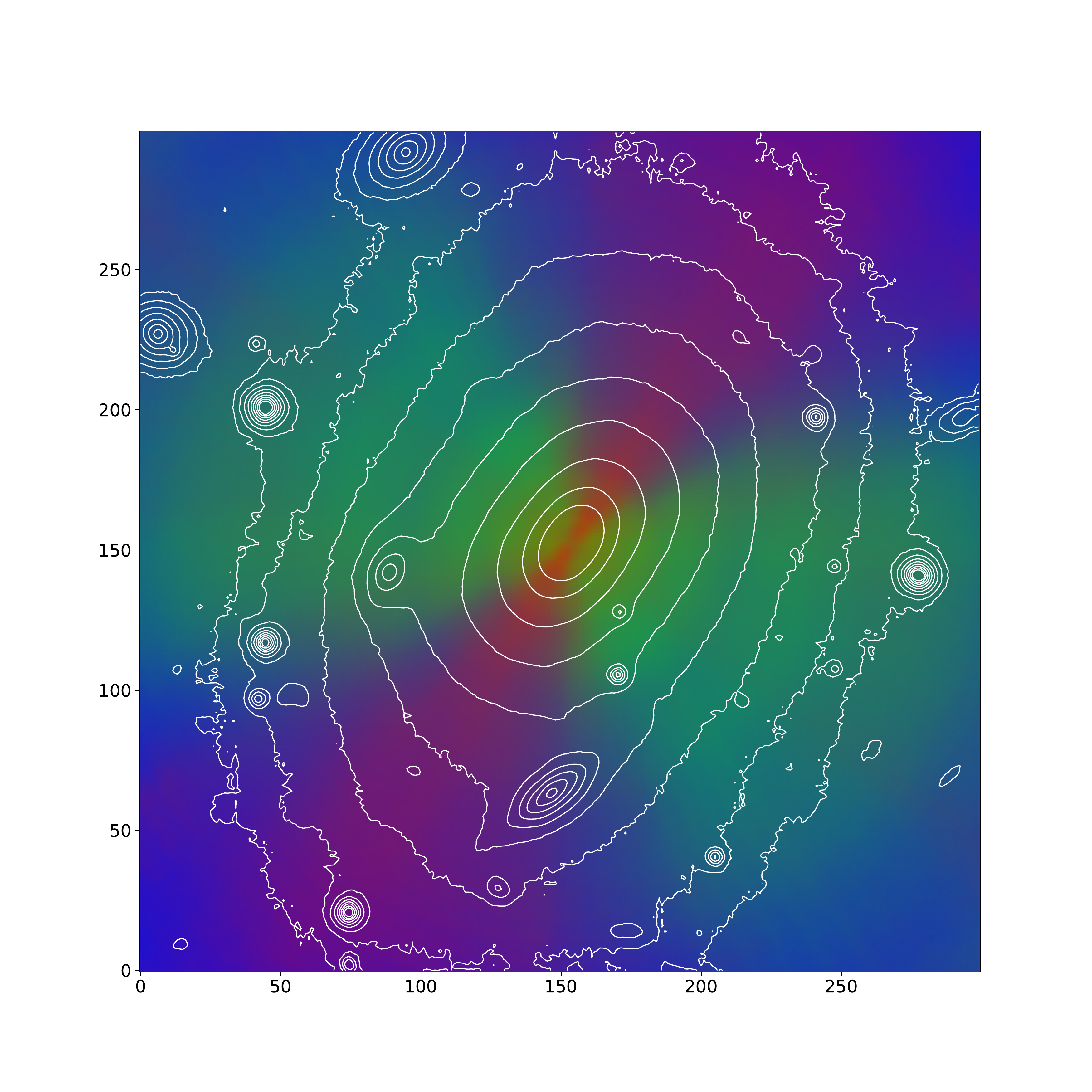} 
  \caption{Orbit distribution colorcoded by different orbit types. Red orbits
    are box orbits, blue orbits are short axis tubes and green orbits are
    long axis tubes.}\label{fig:orbit_dist_rgb}
\end{figure}
In Fig. \ref{fig:orbit_dist_rgb} we show the distribution of the three different orbit types as an RGB image. Perhaps unexpectedly, the weight distribution of short axis tubes dominates over the other two orbit types (in projection) along the minor axis. The reason for this is that the long axis tube orbits in this galaxy are mainly found in the inner part, and the box orbits tend to align along the galaxy's long axis.

\begin{figure*}
   \includegraphics[angle=0, width=.49\textwidth]{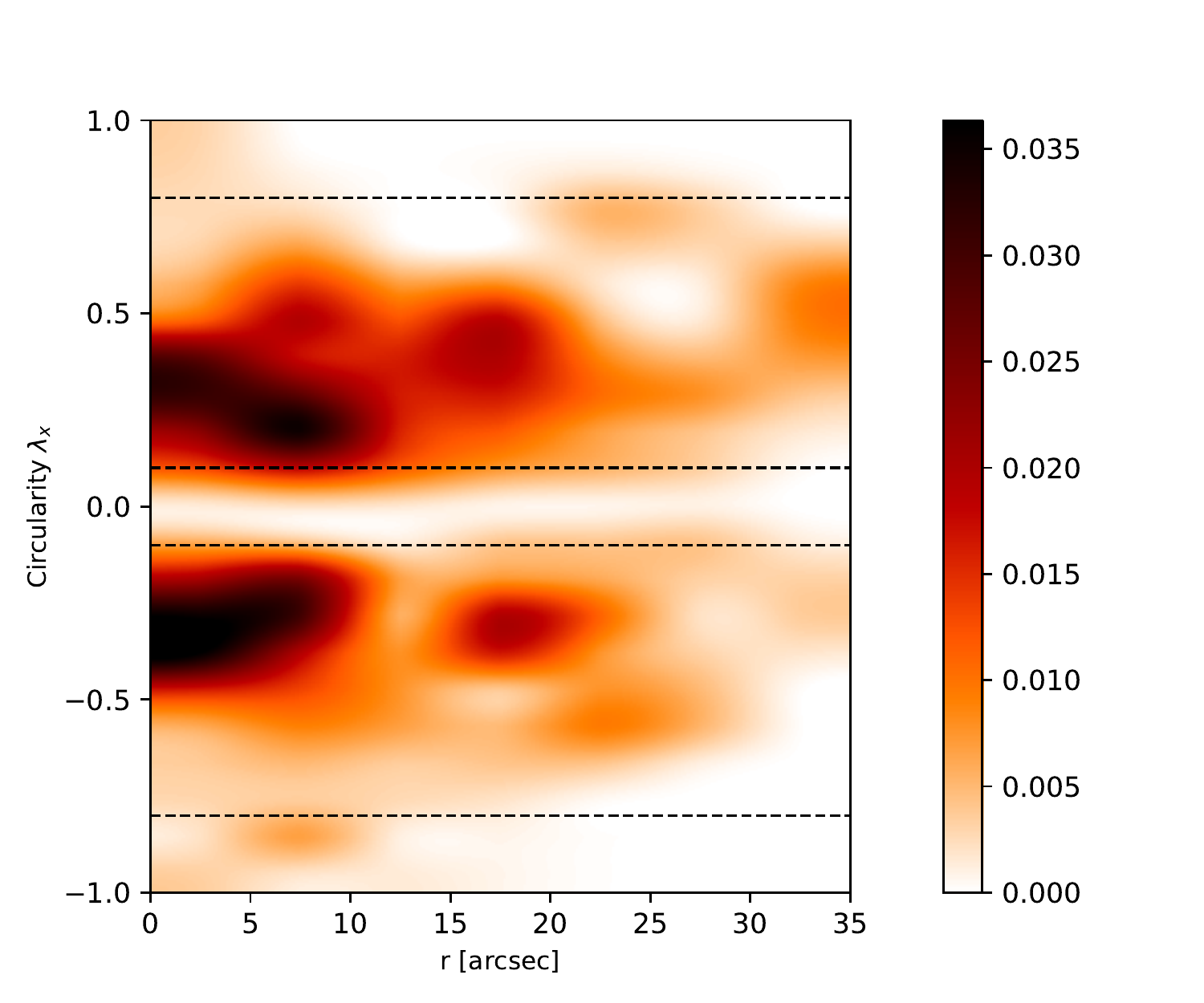} 
  \includegraphics[angle=0, width=.49\textwidth]{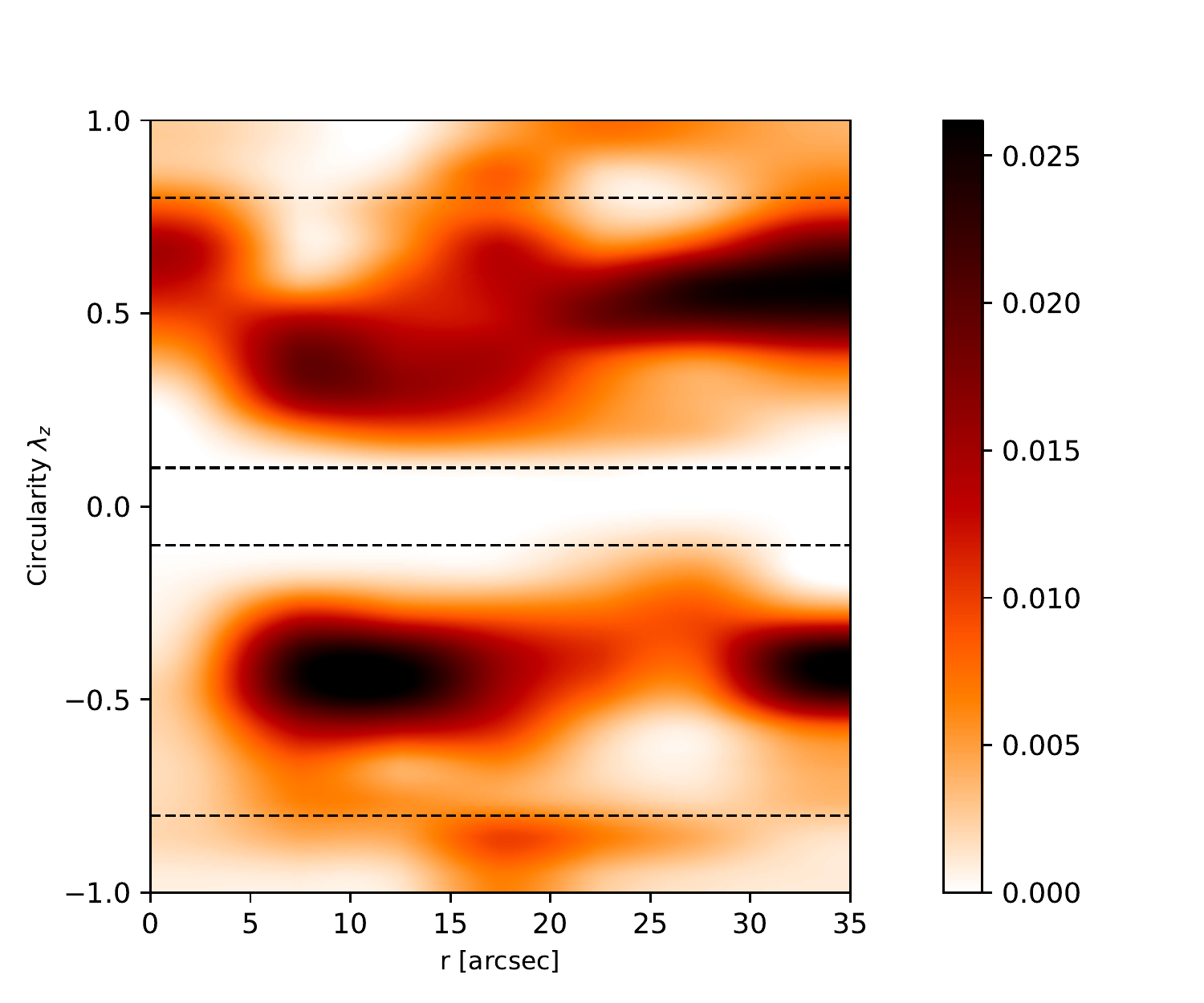} 
  \caption{Average orbital circularity of tube orbits as a function of radius for all models consistent with the best fit model. Darker colours imply a higher density of orbits. The dashed lines separate hot orbits, warm orbits and cold orbits.  }\label{fig:orbit_circ}
\end{figure*}

\begin{figure*}
  \includegraphics[angle=0, width=.99\textwidth]{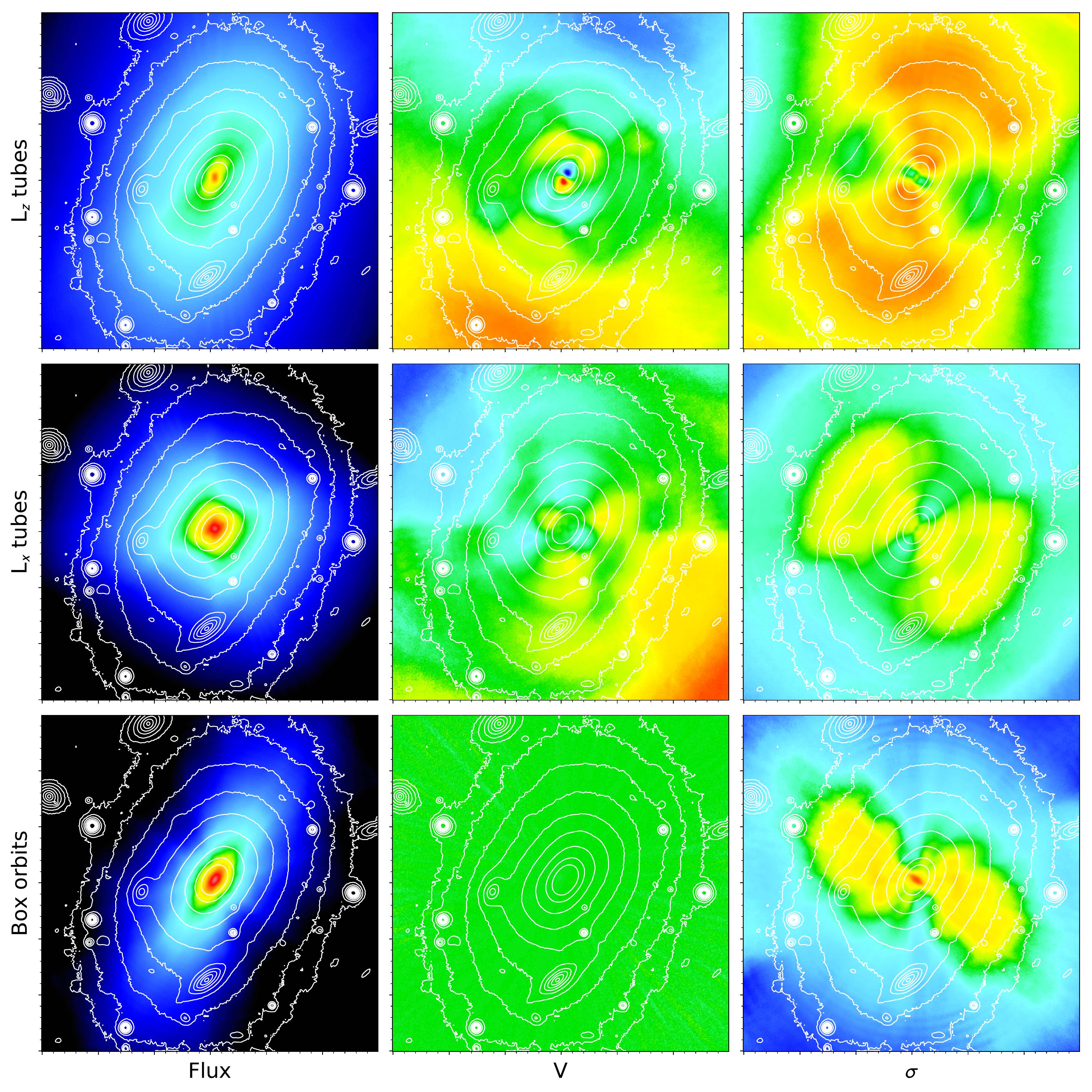} 
  \caption{Kinematics and flux distribution of different orbit types. Left
    panels show the intensity, middle panels the average LOS velocity and
    right panels the LOS velocity dispersion of each orbit type. The upper,
    middle and lower panels are short axis tubes, long axis tubes and box
    orbit.}\label{fig:orbit_dist_kin}
\end{figure*}

To show the orbit distribution in more detail we make diagrams of the circularity phase space, with circularity defined as
\begin{eqnarray}
\lambda_{\bullet} = \frac{\overline{L_{\bullet}}}{r\times\overline{V_c}},
\end{eqnarray}
with $\bullet = x$ or $z$, which has been used by e.g. \citet[][]{BoaWeivan16} to study the orbit distribution in NGC 3998 and \citet{Zhuvanvan18} to study the orbit distribution in CALIFA galaxies. A value of $\lambda$ close to 1 denotes a circular orbit, and values close to zero a radial orbit. The sign of $\lambda$ defines the direction of rotation.

We show the weight distribution of the orbits in Fig. \ref{fig:orbit_circ}, for both long (left panel) and short (right panel) axis tube orbits, averaging over all models (in run 2) that are consistent with the best-fit model. As box orbits have  $\lambda_{x,z} = 0$, we leave those out of both diagrams; we similarly leave out long (short) axis tube orbits from the $\lambda_z$ ($\lambda_x$) diagrams. As expected for massive ellipticals, there is little contribution from cold orbits ($\lambda \approx 1$).

Long axis tube orbits are confined predominantly to the centre of the galaxy, as can also be seen in Fig. \ref{fig:orbit_dist_rgb}. Both prograde and retrograde orbits long axis tubes are present. The observed reversal in angular momentum along the minor axis can be attributed to the somewhat different distributions of the pro- and retrograde orbits. The latter dominate at lower radii over the prograde orbits. The effect of this can be seen also in the middle panel of Fig. \ref{fig:orbit_dist_kin}. Here the middle 3 panels show the distribution, velocity and velocity dispersion of the long axis tubes. The reversal in the kinematics is visible in the centre of the middle panel, and leads to an increase in dispersion seen in the middle right panel, typically also seen in 2$\sigma$ galaxies \citep[galaxies with counterrotating streams, e.g.][]{KraEmsCap11}. This is very similar to the KDC in NGC 4365 \citep{vanvanVer08}.

Contrary to long axis tubes, short axis tubes are found also at larger radii. The prograde orbits are present at all radii, whereas the retrograde orbits, although also present at smaller radii, start dominating at radii of about 10\arcsec.  At larger radii the prograde short axis tubes become dominant again and are responsible for more regular rotation at these radii. The observed two orbital reversals are thus entirely due to the enhanced amount of retrograde short axis tubes with respect to prograde short axis tubes at radii of $\sim$ 10\arcsec.

Many early simulations studying counter-rotating core formation in galaxy centres required gas to be present \citep[e.g.][]{BalQui90,HofCoxDut10}. As central star formation would diminish the central core after merging \citep[][]{BarHer91}, it is doubtful that a gas-rich KDC formation mechanism operated in PGC 046832, which additionaly would probably destroy the high triaxiality in the centre \citep{HofCoxDut10}.

The simulations of \citet{BoiBouEms10,BoiEmsBou11} show that KDCs are naturally produced in major mergers, and are often only ``apparent'', i.e. formed by superposing counter-rotating discs with radially varying mass ratios. This confirmed the picture of \citep{vanvanVer08}, and is similar to what we see in PGC 046832. Although initially thought to require retrograde mergers, \citet{TsaMacvan15} show that this may not be a necessary requirement, as during a retrograde merger reactive forces can change the orbital spin, producing $\sim 2$ kpc scale KDCs. It is unclear if this mechanism can also create the  $\sim 10$ kpc scale KDC in PGC 046832.

The simulations of \citet{RanJohNaa19} provide another way to produce kinematically decoupled components with a mechanism similar to that of \citet{TsaMacvan15}.  In these simulations kinematic reversals naturally arise from the reversal of the orbital directions of the two supermassive black holes of the progenitor galaxies by torques after pericentre passages; these torques result from a combination of the stellar centres not being exactly on the major axis of the merger remant and tidal self friction. In order to get a central kinematically decoupled component in their simulations requires a major dry merger with a mass ratio $q > 1/3$. This scenario is attractive as it naturally explains the two minor axis orbital reversals with a single merger scenario. However, \citet{RanJohNaa19} state that the inner KDC becomes more prominent with increasing black hole mass. Given the poor constraint from the triaxial models it is not clear if the central MBH is actually heavy enough to produce such KDCs and additionally if the mergers described in \citet{RanJohNaa19} that can produce such a KDC would allow a (inner) prolate shape.

\subsection{Chemical tagging}

The complex kinematics of PGC 046832 with multiple angular momentum reversals point at a formation scenario that involves one or more mergers. The fact that stars from different orbital families or with different angular momentum dominate at different locations in the galaxy may provide a means to infer the origins of these stars through their stellar populations. One caveat however is that the recovery of the light fraction in different orbits may not be fully determined by the data \citep[see e.g.][]{NeuThoSag21}. In this Section we associate orbital features with stellar population parameters.

As assigning stellar population parameters directly to individual orbit bundles leads to an underconstrained fit, we follow \citet{PocMcDZhu19} who employed chemical tagging for NGC 3115, and bin orbits together based on their circularity and radius. Contrary to \citet{PocMcDZhu19}, our data do not warrant a detailed separation in circularity. Instead we divide the orbits into 5 different bins based on the orbit type and rotation direction: {\it hot orbits} with $|\lambda_{x}| < 0.1$ and $|\lambda_{z}| < 0.1$, {\it prograde and retrograde long axis tubes} with resp. $\lambda_{x} > 0.1$ or   $\lambda_{x} < -0.1$, and {\it prograde and retrograde short axis tubes} with resp. $\lambda_{z} > 0.1$ or   $\lambda_{z} < -0.1$. These definitions cover all the orbits in the orbit library: we ensured that there are no orbits that simultaneously have  $|\lambda_{x}| >= 0.1$ and $|\lambda_{z}| >= 0.1$.

We divide the radial direction of each of these 5 orbit type bins into 12 radial bins to have as few free parameters as possible without having to worry about internal gradients in bins. Each orbit-type--bin combination has 3 SSP parameters (age,[Z/H] and [$\alpha$/Fe]). This means that we have in total 180 free parameters. We space bins in the inner parts more closely together. As each of the orbits in the library is associated with exactly one radial bin, each of the orbits from the orbit libray has therefore a combination of age,metallicity and $\alpha$-enhancement associated with it. Using MILES models, we convert these values of the SSP parameters to index values for the indices H$\beta$, Fe5015, Fe5270, Fe5335 and Mg b. For each Voronoi bin, we subsequently calculate the light weighted equivalent width of each index. We then calculate a $\chi^2$ by summing the squared difference of the predicted index values and observed index values, divided by the uncertainties.  

As we expect stellar population parameters to vary smoothly with radius, we impose a regularization to maximise smoothness between radial bins of the same orbit type of the form  $\chi^2_{R} = \zeta \sum_i (x_i + x_{i+2}  - 2 x_{i+1})^2$ , which disadvantages local curvature in $\log$(age)/metallicity/[$\alpha$/Fe] in consecutive bins $x_i$. We use the same value of $\zeta$ for the regularization on $\log$(age) and metallicity, but use a value that is 4 times higher for [$\alpha$/Fe], as the parameter range is smaller than that of metallicity and age. This regularization should lead to as smooth as possible radial profiles, without increasing the $\chi^2$. We find that $log_{10}(\zeta) = -1$ is the highest value for which the  $\chi^2$ does not increase significantly. Higher values (e.g. $log_{10}(\zeta) \ge 0 $ ) lead to an increase in $\chi^2 \ge 3$.    

Fitting the stellar populations of the different orbital families thus leads to an inversion problem with 180 free parameters. This conversion is not completely trivial, as the conversion from SSP parameters to index values is non-linear, and the parameters have bounds. To do the inversion, we use the Limited-Memory Broyden–Fletcher–Goldfarb–Shanno algorithm, which is a modification of the Broyden–Fletcher–Goldfarb–Shanno algorithm particularly well-suited for a high-dimensional parameter space \citep{ByrLuNoc95}. Once the best-fit solution is determined, we calculate uncertainty intervals from the diagonal elements of the inverted Hessian.

\begin{figure*}
   \includegraphics[angle=0, width=.9\textwidth]{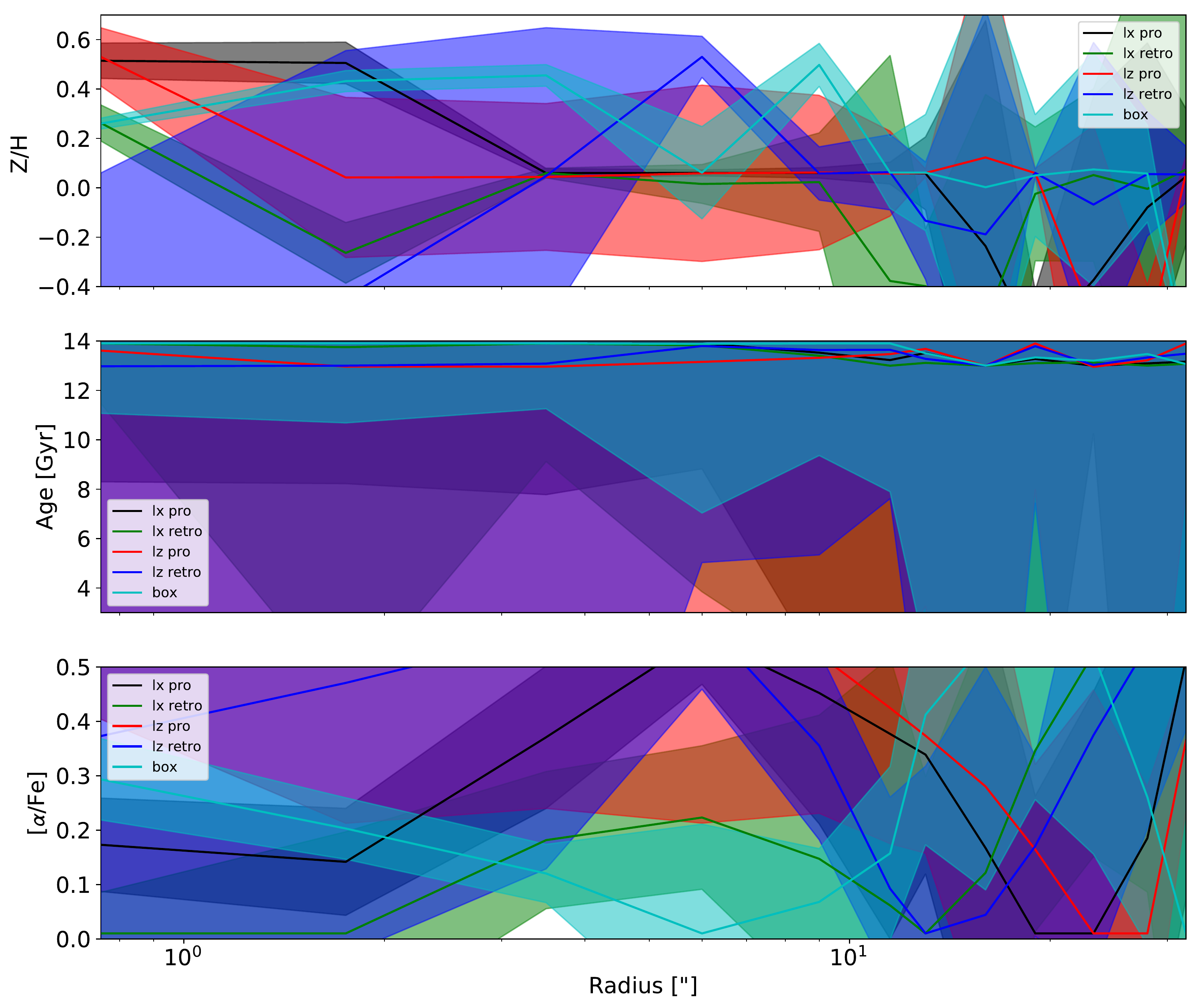}
  \caption{The radial distribution of age, metallicity and [$\alpha$/Fe] of different orbit types, with standard uncertainties based on the inversion of the Hessian. Orbits of all types exhibit old ages. Metallicity profiles show no indication for different chemical origins near the centre of the galaxy. }\label{fig:orbittag}
\end{figure*}
 
Fig. \ref{fig:orbittag} shows the distribution of SSP parameters of the different orbit families. Even with regularization, we see no clear difference in age, metallicity or [$\alpha$/Fe] for different orbital families. We conclude therefore that there is no evidence from the stellar population parameters that prograde and counter-rotating orbits originate from different progenitor galaxies.

\section{Conclusions}
We present VLT/MUSE data of PGC 046832, one of the massive BCGs in the Shapley super cluster. The MUSE data allow us to extract spatially resolved kinematics and stellar populations. We fit the kinematics with a LOSVD consisting of a Gaussian with Gauss-Hermite moments up to $h_6$. The kinematics of PGC 046832 are complex, and show two reversals of line-of-sight velocity along the projected major axis of the galaxy, and a reversal along the minor axis.

Using archival HST data and VLT/OmegaCAM data we model the photometry as a sum of a core-S\'ersic model and a S\'ersic model. We confirm the presence of a core in this galaxy with size $R_b =0\farcs41$ and a strong isophotal twist in the photometry.

The stellar populations show a central old (13 Gyr) and metal rich (Z=0.3) population, with a negative gradient in metallicity (and age) toward the outskirts. Using synthesized stellar population models with varying abundances, we find a moderate enhancement ([$\alpha$/Fe] $\sim$ 0.2) with no clear radial gradient.

We model the kinematics and surface brightness distributions of the galaxy simultaneously using Schwarzschild's orbit superposition method under the assumption of triaxiality  with the code of \citet{vanvan09}. The many kinematic reversals and the isophotal twist allow us to constrain the viewing angles of this galaxy to high accuracy. These angles suggest a highly triaxial shape in the centre of the galaxy ($T$ close to 1), where $p \sim q$, and a very low triaxiality ($T\sim0.2$) in the outer parts of the galaxy.

We study the orbit distribution in the best-fit Schwarzschild model. Most tube orbits have a low circularity, as expected for orbits in a massive elliptical galaxy. Long axis tube orbits with positive circularity extend slightly farther beyond the centre than their negative counter parts, but are less prominent in the very centre. This behaviour is responsible for the reversal of the velocity sign along the projected minor axis. Short axis tubes are present at all radii. The kinematic reversals along the major axis are caused by stars on retrograde rotating orbits which are predominant at 10\arcsec from the centre. Using chemical tagging of orbits, we find no evidence that stars on prograde or retrograde short axis or long axis tube orbits have a different chemistry.

Using axisymmetric Jeans models, axisymmetric Schwarzschild models and the triaxial Schwarzschild models, we infer the mass of the central supermassive black holes. The Jeans models provide the highest mass (log M$_{BH}$ = 9.98$\pm$ 0.02). The axisymmetric code suggests a lower mass (log M$_{BH}$ = 9.8$\pm$ 0.1), whereas the triaxial code provides only an upper limit to the black hole mass. We attribute these differences to the absence of orbital families in axisymmetric models.

Although it is possible that the velocity reversals in the galaxy result from a merging event involving two very massive black holes, similarly to what has been seen in the simulations of \citet{RanJohNaa19}, the low black hole mass implied by the triaxial Schwarzschild models makes this scenario unlikely. The nearly prolate inner shape of the galaxy is most likely the result of a major merger. Contrary to NGC 3115, for which chemical tagging has shown different stellar populations associated with specific orbits, we do not find evidence in PGC 046832 for this. This either means that the Schwarzschild model does not perfectly recover the light fractions in different orbits, or that the stellar populations are too mixed to be distinguishable. The lower metallicity and more regular rotation in the outer parts of the galaxy are consistent with both in situ formation and ex situ accretion.

\section*{Acknowledgements}
MdB and DK acknowledge financial support through the grant GZ: KR 4548/2-1 of the Deutsche Forschungsgemeinschaft. JB acknowledges support by Funda\c{c}{\~a}o para a Ci\^encia e a Tecnologia (FCT) through the research grants UID/FIS/04434/2019, UIDB/04434/2020, UIDP/04434/2020. Based on observations made with the NASA/ESA Hubble Space Telescope, and obtained from the Hubble Legacy Archive, which is a collaboration between the Space Telescope Science Institute (STScI/NASA), the Space Telescope European Coordinating Facility (ST-ECF/ESA) and the Canadian Astronomy Data Centre (CADC/NRC/CSA). Based on observations made with ESO Telescopes at the La Silla Paranal Observatory under programme IDs 095.B-0127, 096.B-0061, 098.B-0240 and 102.B-0327.

\section*{Data Availability}

Raw MUSE data are available in the ESO archive. Kinematics are available from the author on reasonable request.



\bibliographystyle{mnras}
\bibliography{paper} 



\section*{Supporting information}
Supplementary figures are available at MNRAS online.\\
Figures C1-C4\\
Figures D1-D2

\appendix

\section{Second moment of galaxies with non-zero GH coefficients}\label{app:gh}
The velocity distribution function $f$ along any line of sight can be written as a Gauss-Hermite series with $h_0=1$ and $h_1 = h_2 = 0$ \citep{vanFra93}:
\begin{eqnarray}
f(y) = \gamma\frac{e^{-\frac{y^2}{2}}}{\sqrt{2\pi}\sigma}\left(1+ \sum_{i=3}^{\infty} h_i H_i(y)\right),
\end{eqnarray}
with $y=\frac{v-V}{\sigma}$ and 
\begin{eqnarray}
  H_3(y)=(4 y^3 - 6 y)/\sqrt{12}\\
  H_4(y)=(4y^4 - 12y^2 + 3)/\sqrt{24}\\
  H_5(y)=(4y^5 - 20y^3 +15y)/\sqrt{60}\\
  H_6(y)=(8y^6 - 60y^4 + 90y^2 -15)/\sqrt{720}.
\end{eqnarray}
The second moment can thus be written as:
\begin{eqnarray}
 &\int f(v) v^2 dv = \int f(y) (y+V/\sigma)^2 \sigma^3 dy =\nonumber\\
& \gamma\sigma^2\int  \frac{e^{-\frac{y^2}{2}}}{\sqrt{2\pi}}(y+V/\sigma)^2 \left( 1 +  h_3 H_3(y) + h_4 H_4(y) + \ldots\right) dy\nonumber\\
 &=\gamma( V^2 + \sigma^2 + 2\sqrt{3} h_3 V\sigma + \frac{h_4}{\sqrt{24}}\left(27\sigma^2 + 3 V^2 \right)\nonumber \\
 &+ \sqrt{15} h_5 \sigma V + \frac{h_6}{\sqrt{720}}\left( 195\sigma^2 + 15 V^2\right) + \ldots ),
  \end{eqnarray}
with normalization
\begin{eqnarray}
\gamma = \left(1 + h_4 \frac{3}{\sqrt{24}} + h_6 \frac{15}{\sqrt{720}} + \ldots \nonumber \right)^{-1}.
  \end{eqnarray}

\section{1-dimensional SSP profiles}\label{apx:ssp}
\begin{figure}
  \includegraphics[angle=0, width=.49\textwidth]{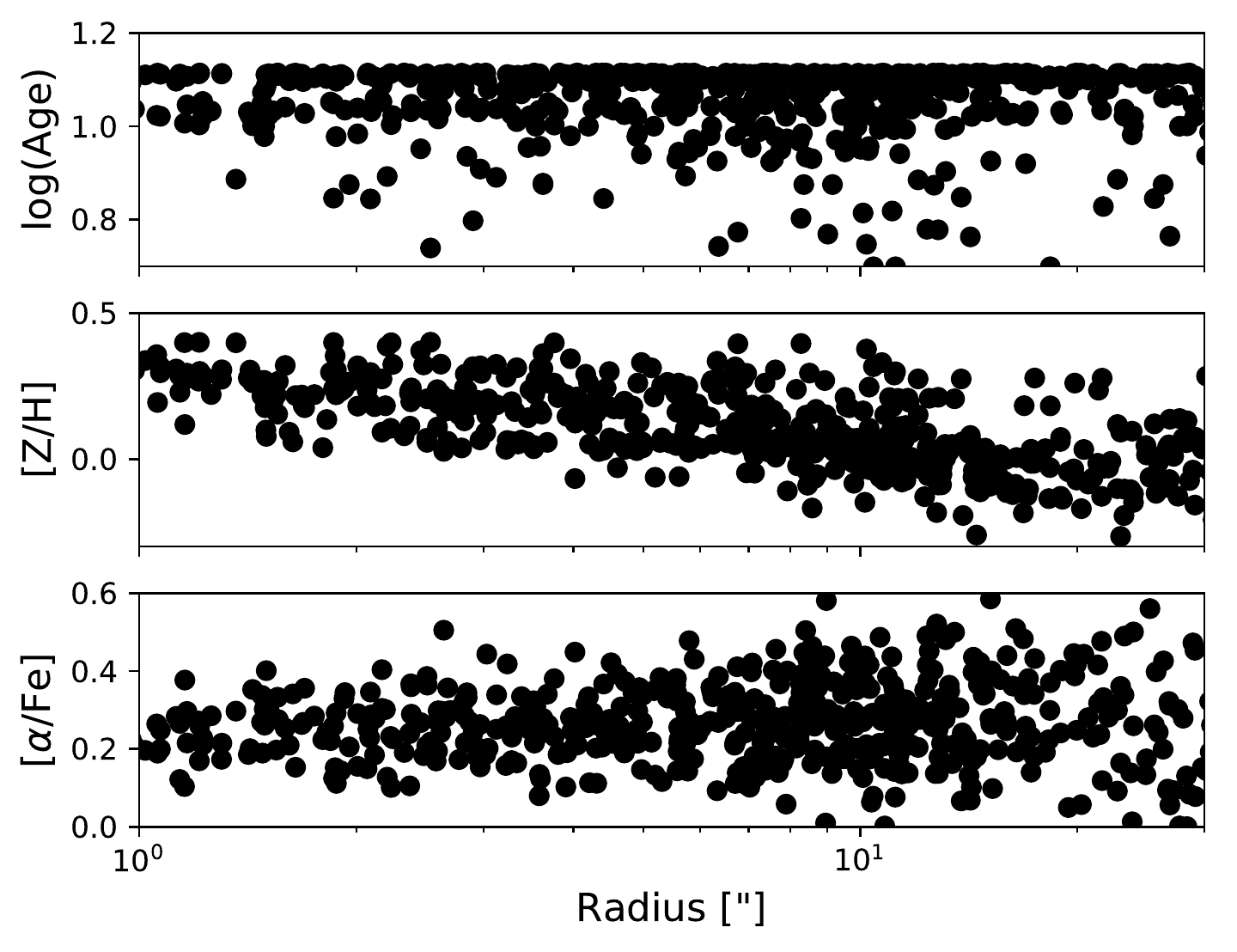}
  \caption{Mass-weighted SSP parameters as a function of elliptical major axis distance to the galaxy centre.}\label{fig:ssp_1d}
\end{figure}

In Fig. \ref{fig:ssp_1d} we show the inferred mass-weighted SSP parameters as a function of distance to the galaxy centre. 

\section{Axisymmetric MGE and model fits}\label{apx:axi}
\begin{table}
\caption{Components of the axially symmetric MGE used for the Jeans models and
  axisymmetric Schwarzschild models.}
\label{tab:MGE_notwists}  
\begin{tabular}{cccc} 
\hline
No. & SB$_0$  & $\sigma$  & $q'$ \\
 & [L$_{\odot}$/pc$^2$] & [arcsec]  & \\
\hline
1&  2139.9 &   0.084 &    0.80 \\  
2& 3793.0 &  0.26  & 0.91 \\  
3& 7924.0 &  0.48  & 0.68\\   
4& 2905.2 &  0.95  & 0.62 \\  
5& 1930.1 &  1.74  & 0.66 \\  
6& 481.1 &  3.59  & 0.62 \\  
7& 367.2 &  6.05  & 0.72 \\  
8& 114.6  & 12.44  & 0.65 \\  
9& 57.0  &  20.91 &  0.83 \\  
\end{tabular}
\end{table}
\begin{figure*}
  \includegraphics[angle=0, width=.9\textwidth]{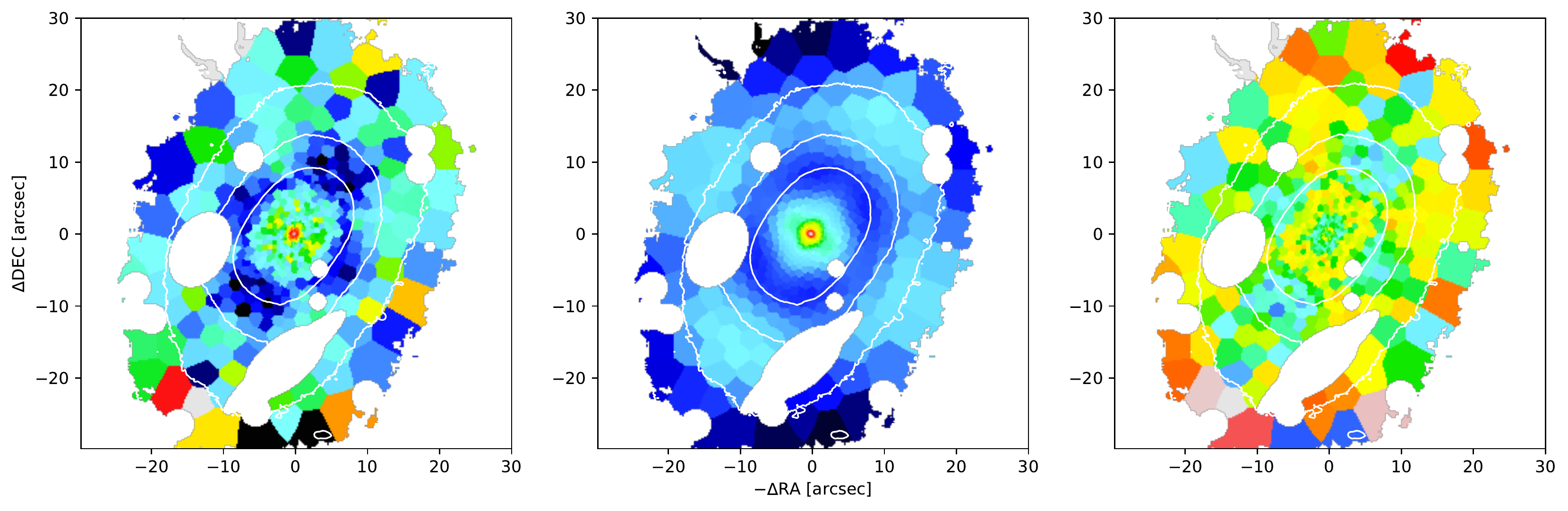} 
  \caption{Left panel: observed second moment ($V_{\mathrm{RMS}}$). Middle panel: best-fit axisymmetric JAM model. Right panel: residuals after model subtraction. Model and data are plotted on a scale of 270-346 km/s, residuals are plotted between -50 and 50 km/s. }\label{fig:jam_res_oblate}
\end{figure*}
\begin{figure}
  \includegraphics[angle=0, width=.49\textwidth]{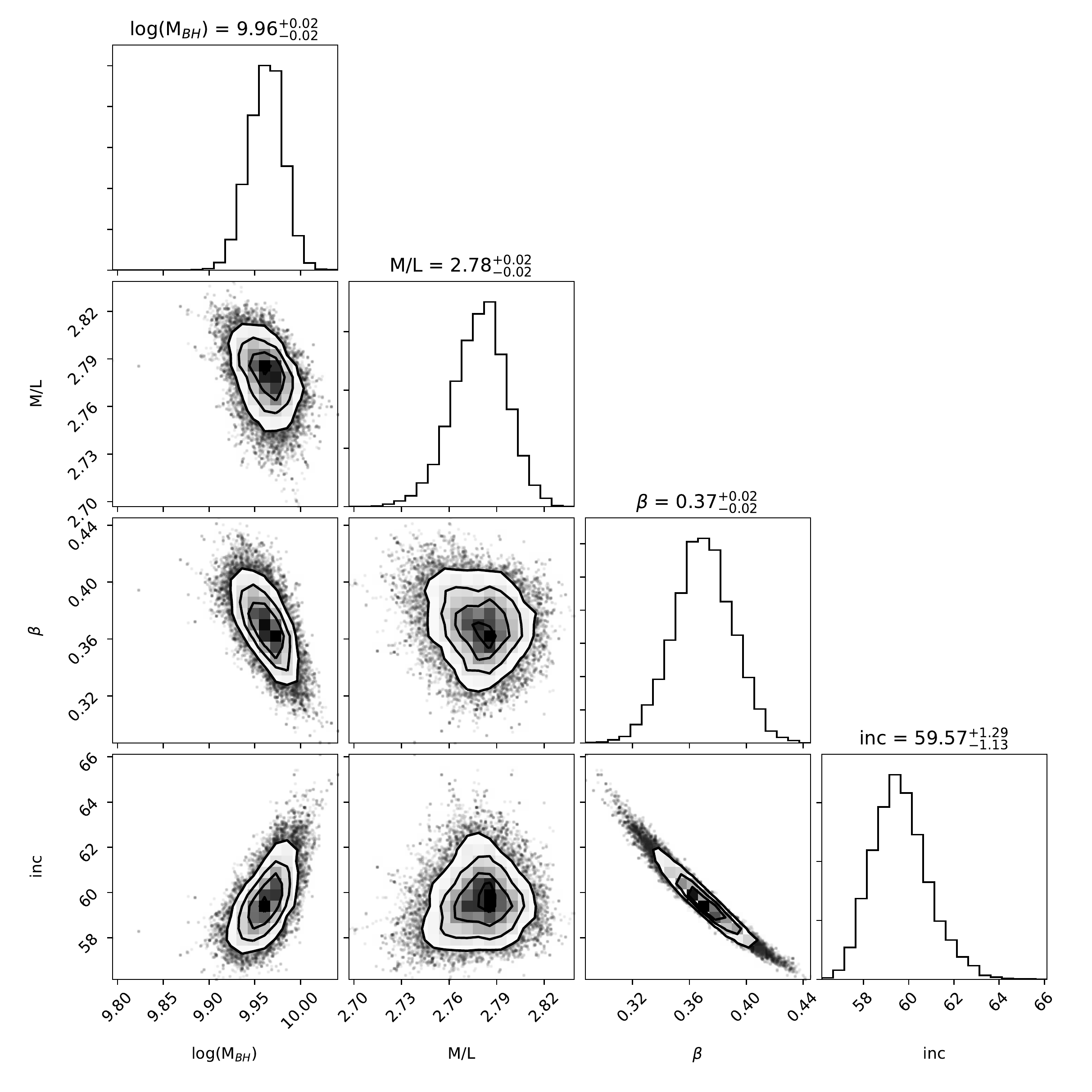} 
  \caption{Corner plot showing the projected distributions of different parameter pairs of the axisymmetric JAM models as well as parameter histograms. The quoted uncertainties in the plot are 1-sigma uncertainties. }\label{fig:jam_mcmc}
\end{figure}
\begin{figure*}
  \includegraphics[angle=0, width=.9\textwidth]{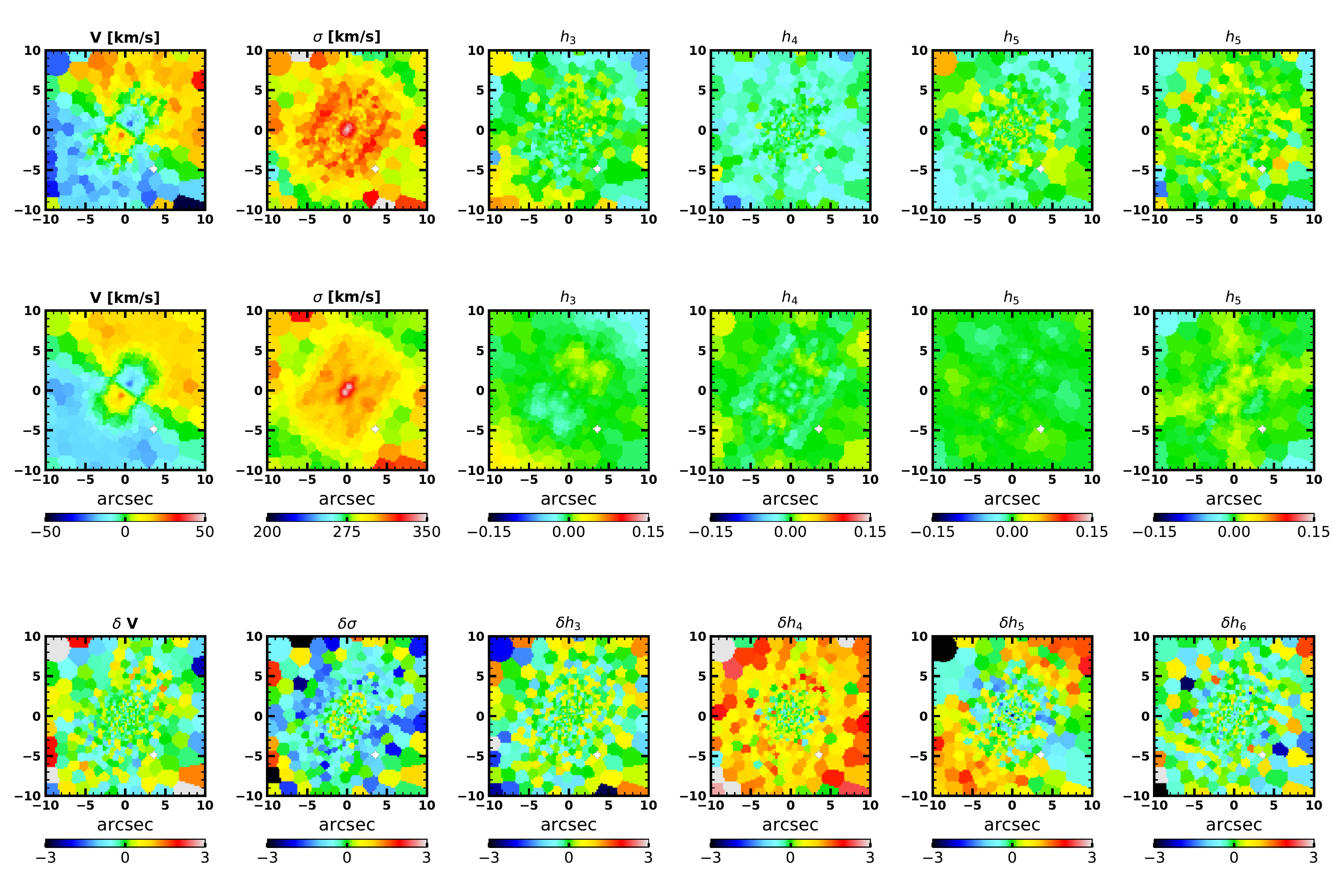}
  \caption{Data and model comparison for the best-fit axisymmetric Schwarzschild model. First row shows the symmetrized kinematic maps. The second row the kinematic maps of the best-fit Schwarzschild model. Residuals normalised by the uncertainties are shown in the third row. }\label{fig:axi_schw_res}
  \end{figure*}
  \begin{figure*} 
  \includegraphics[angle=0, width=.9\textwidth]{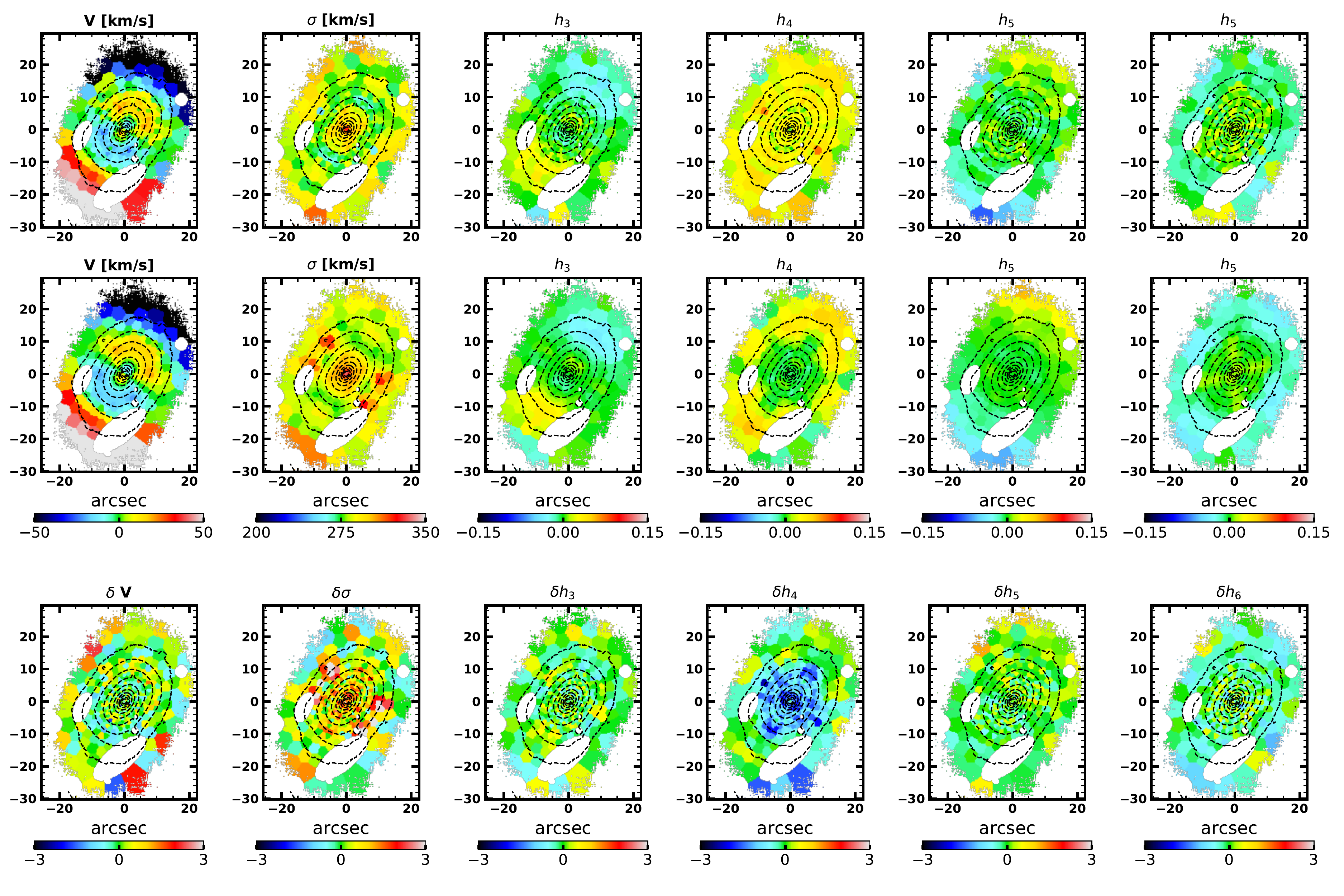} 
  \caption{Data and model comparison for the best-fit axisymmetric Schwarzschild model of the full MUSE field-of-view. Layout as in Fig. \ref{fig:axi_schw_res}  }\label{fig:axi_schw_res2}
\end{figure*}

In Table \ref{tab:MGE_notwists} we present the MGE used in the axisymmetric models. Fig. \ref{fig:jam_res_oblate} shows the best fit oblate Jeans model. Fig. \ref{fig:axi_schw_res} shows the model data comparison of the best-fit axisymmetric Schwarzschild model.

\section{BH mass fits including the Ca{\rm II} triplet.}\label{sec:cat}
\begin{figure*} 
     \includegraphics[angle=0, width=.9\textwidth]{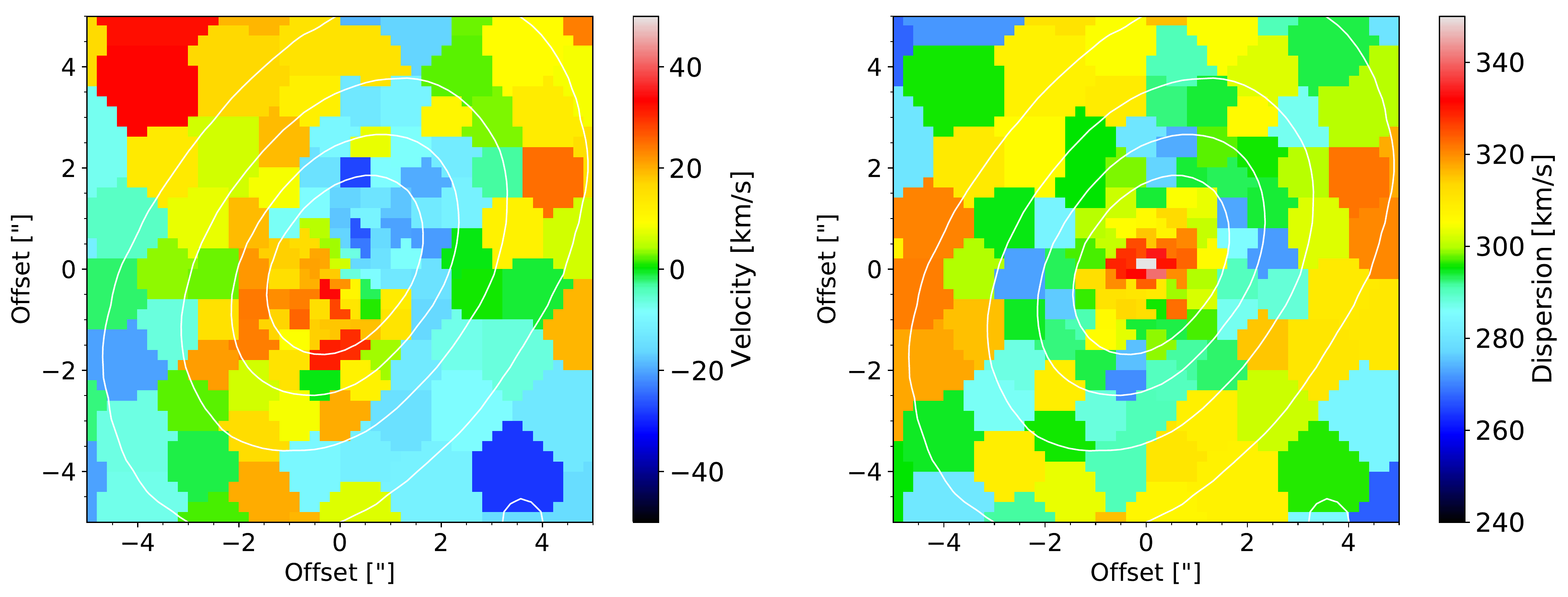}
     \includegraphics[angle=0, width=.9\textwidth]{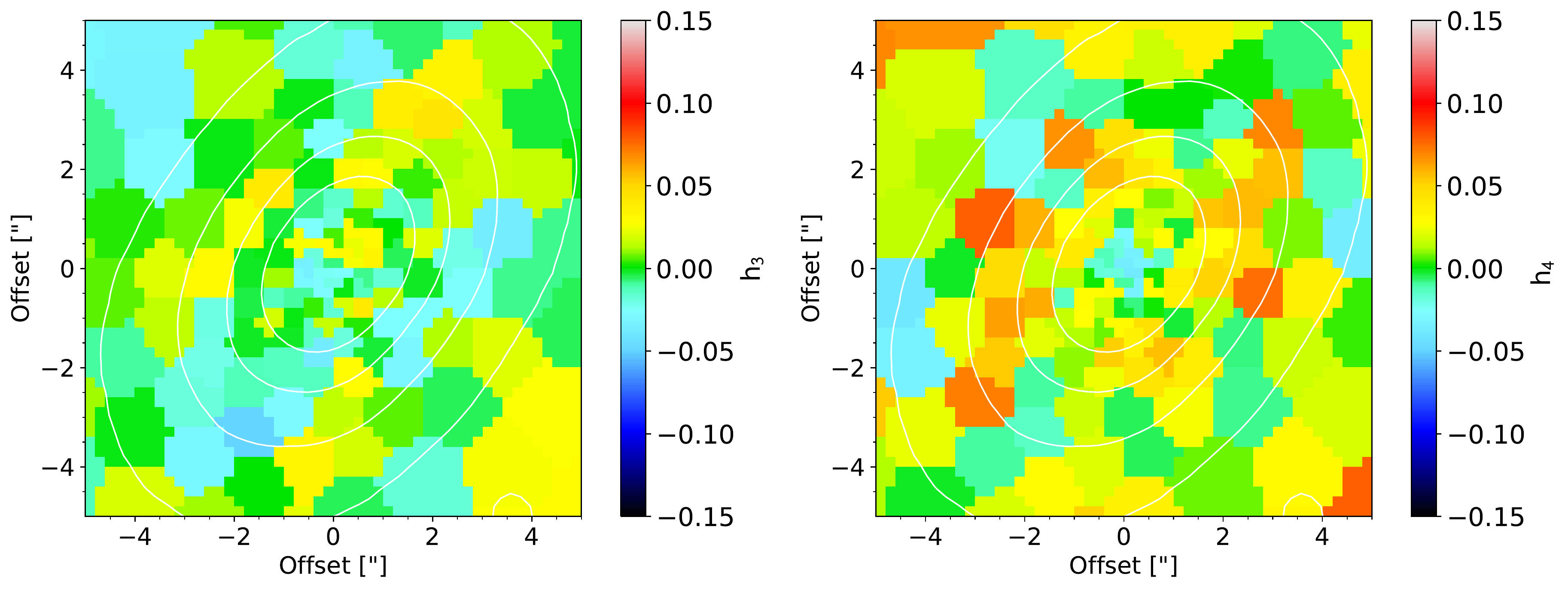}
  \caption{Central kinematics derived from the spectral region around the Calcium triplet. Upper panels show line-of-sight velocity and dispersion, lower panels the first two Gauss-Hermite moments.}\label{fig:kin_cat_maps}
\end{figure*}

The spatial resolution of the observations is higher in the NIR than in the blue part of the spectrum. We find that the PSF of the cube in the wavelength range 7900-9300\AA\ has a FWHM of 0\farcs49. To optimally benefit from this increased resolution we extract kinematics up to $h_4$ to the spectral region around the Calcium triplet. This region is however severly affected by telluric aborption and not fully subtracted sky lines. We found however that using a combination of telluric correction with \textsc{molecfit} \citep{SmeSanNol15} and sky line correction with \textsc{zap} with a narrow continuum filter width (\texttt{cfwidth=10}) leads to significantly improved spectra which can be used for extracting kinematics. It is however only possible to extract reliable kinematics in the very brightest part of the galaxy, and we therefore restrict the fits of the dynamical models to data within 5\arcsec.
The derivation of the kinematics proceeds along the same lines as for the blue part of the spectrum. We tesselate the data to achieve S/N of 50/\AA\ per Voronoi bin, which we subsequently fit with \textsc{pPXF}.
We use the E-MILES synthesized single stellar population spectra as input templates for \textsc{pPXF}. We tried deriving the kinematics also with stellar libraries, but for not fully understood reasons, this led to a significantly low-biased velocity dispersion and small biases in the Gauss-Hermite terms. 
Using the E-MILES library we do still see a very small bias in $h_3$. We show the kinematic maps in Fig. \ref{fig:kin_cat_maps}.
Fitting Jeans models to the central kinematics leads to a slightly higher black hole mass ($\log{M_{\mathrm{BH}}}=10\pm0.05$) and a slightly lower M/L ($\Gamma = 2.5 \pm 0.1$). To fit Schwarzschild models to the these data, we include the outer kinematics derived from the blue part of the spectrum in the fit. We disable the kinematics in the blue within the central 5\arcsec by increasing their error bars by a factor 3. Two sets of orbit libraries are then generated simultaneously for the high and low-resolution data.  
We note that despite the increase in spatial resolution, the fits do still not provide a detection and only give an upper limit ($\log{M_{\mathrm{BH}}} < 9.3$)
 on the black hole mass to a detection.

\begin{figure*} 
    
     \includegraphics[angle=0, width=.9\textwidth]{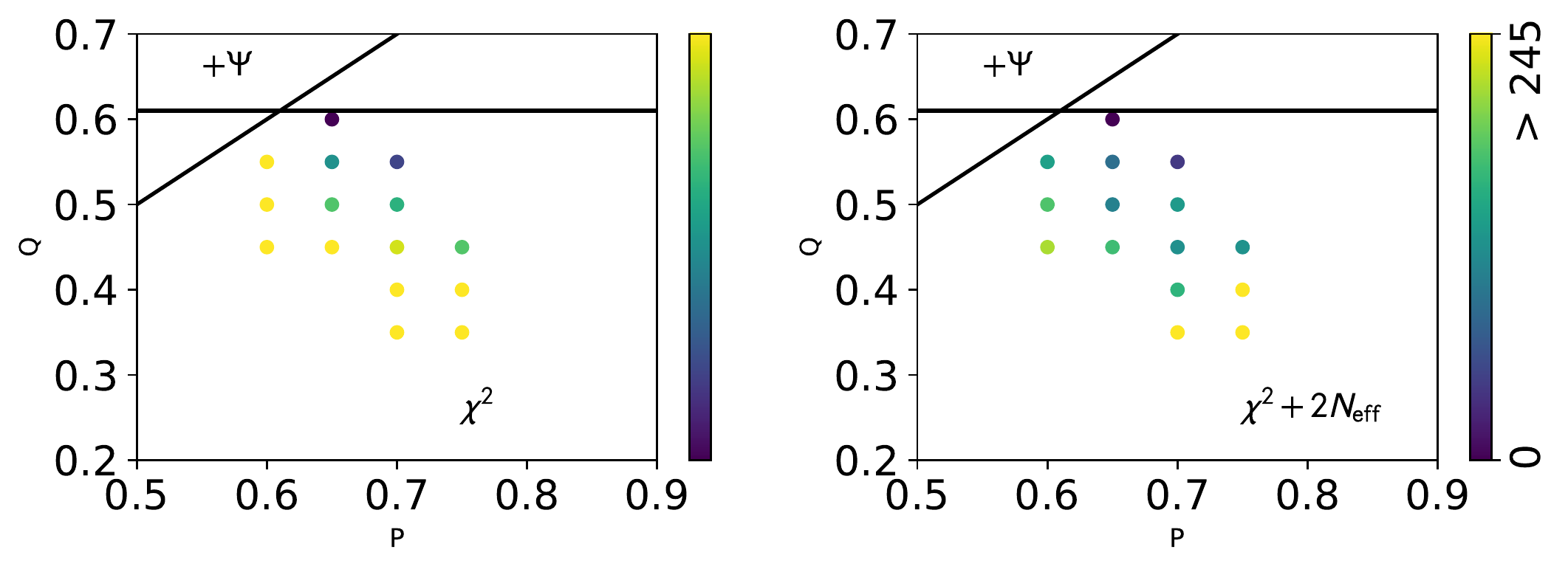}
  \caption{ Comparison between $\chi^2$ of the 21 best fitting models from the grid search (left) and after correcting for the effective number of free parameters (right) using the methodology of \citet{LipTho21}.}\label {fig:chi2_lipka}
\end{figure*}


\bsp	
\label{lastpage}
\end{document}